\documentclass[12pt,a4paper]{article}
\usepackage{amsmath}

\usepackage{fancyhdr}
\renewcommand{\maketitle}{
    \begin{center}
      \Large
        {\bf Abraham-Lorentz-Dirac Equation \\ in 5D Stuekelberg Electrodynamics}
        \vskip .3 true cm
      \small
        Martin Land \\
        \vskip .3 true cm
        Department of Computer Science \\
        Hadassah College \\
        37 HaNevi'im Street, Jerusalem \\
email: martin@hadassah.ac.il
      \end{center}
      \vskip .5 true cm
}
\input{tcilatex}
\begin{document}

\title{}
\author{}
\maketitle

\begin{abstract}
We derive the Abraham-Lorentz-Dirac (ALD) equation in the framework of the
electrodynamic theory associated with Stueckelberg manifestly covariant
canonical mechanics. In this framework, a particle worldline is traced out
through the evolution of an event $x^\mu\left(\tau\right)$. By admitting
unconstrained commutation relations between the positions and velocities,
the associated electromagnetic gauge fields are in general dependent on the
parameter $\tau$, which plays the role of time in Newtonian mechanics.
Standard Maxwell theory emerges from this system as a $\tau$-independent
equilibrium limit. In this paper, we calculate the $\tau$-dependent field
induced by an arbitrarily evolving event, and study the long-range radiation
part, which is seen to be an on-shell plane wave of the Maxwell type.
Following Dirac's method, we obtain an expression for the finite part of the
self-interaction, which leads to the ALD equation that generalizes the
Lorentz force. This third-order differential equation is then converted to
an integro-differential equation, identical to the standard Maxwell
expression, except for the $\tau$-dependence of the field. By studying this $%
\tau$-dependence in detail, we show that field can be removed from the
integration, so that the Lorentz force depends only on the instantaneous
external field and an integral over dynamical variables of the event
evolution. In this form, pre-acceleration of the event by future values of
the field is not present.
\end{abstract}

\baselineskip7mm \parindent=0cm \parskip=10pt

\section{Introduction}

In classical electrodynamics, fields accelerate charges through the Lorentz
force%
\begin{equation}
m\ddot{z}^{\mu }=eF^{\mu \nu }\left( z\right) \dot{z}_{\nu }
\label{Lorentz-std}
\end{equation}%
and current densities defined on charge worldlines%
\begin{equation}
J^{\mu }\left( x\right) =\int d\tau ~\dot{z}^{\mu }\left( \tau \right)
\delta ^{4}\left[ x-z\left( \tau \right) \right]
\end{equation}%
induce fields through Maxwell's equations%
\begin{equation}
\partial _{\nu }F^{\mu \nu }\left( x\right) =eJ^{\mu }\left( x\right) %
\mbox{\qquad}\mbox{\qquad}\partial _{\lbrack \lambda }F_{\mu \nu ]}\left(
x\right) =0~.
\end{equation}%
In general configurations, multiple charges interact through a complex
system of mutual interactions, but as a first approximation to the motion of
a single charge, the field may be treated as fixed, neglecting both its
evolution under recoil of the sources producing it and the reaction field
induced by the accelerating charge itself. The manifestly covariant system
(1) -- (3) employs an invariant evolution parameter $\tau $ often associated
with proper time; this association is at best inexact, ignoring the
diversity of proper times among the many particles as well as the presumed $%
\tau $-reparameterization invariance of the system.

The Abraham-Lorentz-Dirac (ALD) equation \cite{dirac,barut}%
\begin{equation}
m\ddot{z}^{\mu }=eF_{ext}^{\mu \nu }\left( z\right) \dot{z}_{\nu }+\frac{2}{3%
}e^{2}\left( \dddot{z}^{\mu }-\dot{z}^{\mu }\ddot{z}^{2}\right)  \label{ALD}
\end{equation}%
improves the fixed field approximation, still treating the field $%
F_{ext}^{\mu \nu }$ as external to the dynamics, but incorporating a generic
Green's function solution%
\begin{equation}
F_{rad}^{\mu \nu }\left( x\right) =\partial ^{\mu }A^{\nu }-\partial ^{\nu
}A^{\mu }\mbox{\qquad}\mbox{\qquad}A^{\nu }\left( x\right) =-e\int
d^{4}x^{\prime }~D_{rad}\left( x-x^{\prime }\right) J^{\nu }\left( x^{\prime
}\right)
\end{equation}%
to the Maxwell equations to account for the transfer of energy-momentum from
the charge to the field. Thus, ALD provides an effective one-particle theory
for a charge in an external electromagnetic field, but being a third order
differential equation, admits run-away solutions of the type%
\begin{equation}
\dot{z}^{0}=\cosh \left[ \tau \left( e^{\tau /\tau _{0}}-1\right) \right] %
\mbox{\qquad}\mbox{\qquad}\dot{z}^{1}=\sinh \left[ \tau \left( e^{\tau /\tau
_{0}}-1\right) \right]
\end{equation}%
for the free particle case $F_{ext}^{\mu \nu }=0$, where%
\begin{equation}
\tau _{0}=\frac{2}{3}\frac{e^{2}}{4\pi m}\sim 10^{-24}~\sec  \label{t0}
\end{equation}%
is the time scale associated with a photon crossing a classical electron
radius. This unphysical solution may be eliminated under the assumption of
bounded acceleration, by converting (\ref{ALD}) to the integro-differential
equation%
\begin{equation}
m\ddot{x}^{\mu }\left( \tau \right) =\int_{\tau }^{\infty }ds~\frac{1}{\tau
_{0}}e^{-\left( s-\tau \right) /\tau _{0}}\left[ \mbox{\rule{0cm}{10.5pt}}%
eF_{ext}^{\mu \nu }\left( x\left( s\right) \mbox{\rule{0cm}{10.5pt}}\right) 
\dot{x}_{\nu }\left( s\right) -\tau _{0}m\ddot{x}^{2}\left( s\right) \dot{x}%
^{\mu }\left( s\right) \right] .  \label{ALDI}
\end{equation}%
Although (\ref{ALDI}) suppresses run-away solutions, its causal behavior
presents a new difficulty in the form of pre-acceleration. If the external
field is a constant switched on at $\tau =0$, so that 
\begin{equation}
\frac{e}{\tau _{0}}F_{ext}^{\mu \nu }\left( x\left( \tau \right) %
\mbox{\rule{0cm}{10.5pt}}\right) \dot{x}_{\nu }\left( \tau \right)
\rightarrow \frac{1}{\tau _{0}}F^{\mu }\left( \tau \right) =\frac{1}{\tau
_{0}}F_{0}^{\mu }~\theta \left( \tau \right) ,  \label{causal}
\end{equation}%
then one obtains a non-zero pre-acceleration 
\begin{equation}
m\ddot{x}^{\mu }\left( \tau \right) \simeq \frac{1}{\tau _{0}}F_{0}^{\mu
}\int_{\tau }^{\infty }ds~e^{-\left( s-\tau \right) /\tau _{0}}\theta \left(
s\right) =\left\{ 
\begin{array}{lll}
F_{0}^{\mu }e^{-\left\vert \tau \right\vert /\tau _{0}} & ,%
\mbox{\rule[-0.5cm]{0cm}{1.0cm}} & \tau <0 \\ 
F_{0}^{\mu } & , & \tau >0%
\end{array}%
\right.
\end{equation}%
effective over a period on the order of $\tau _{0}$ \emph{before} the field
is applied. It may be argued that on this time scale, problems of classical
microcausality are overtaken by quantum effects, empirically and
theoretically. Nevertheless, a violation of classical causality posed in
classical electrodynamics seems to raise questions about the interpretation
of the underlying framework, beyond issues of approximation and measurement.

In this paper, we develop the Abraham-Lorentz-Dirac equation in a
formulation of classical electrodynamics associated with a manifestly
covariant canonical mechanics of interacting spacetime events. This
approach, first suggested by Stueckelberg \cite{Stueckelberg} in modeling
pair creation/annihilation, regards the charged event $x^{\mu }\left( \tau
\right) $ as the source of an electromagnetic field, so that event currents
and fields are defined locally at a given spacetime point $x^{\mu }$ and
given moment $\tau $. Microscopic details of event dynamics may differ from
the usual approach, but standard Maxwell theory emerges as the $\tau $%
-static equilibrium limit. Because the relationships among events,
particles, currents, and fields are determined at a finer resolution, this
formalism leads to a generalization of (\ref{ALDI}) for which the causal
structure can be given a natural and consistent interpretation. In the next
section, we review the essential features of Stueckelberg electrodynamics,
then characterize the induced radiation field in section 3, and derive the
ALD equation for the Stueckelberg framework in section 4.

\section{Stueckelberg Off-Shell Electrodynamics}

In seeking a classical description of pair creation/annihilation as a single
worldline generated dynamically by the evolution of an event $x^{\mu }\left(
\tau \right) $, Stueckelberg proposed \cite{Stueckelberg} a generalized
Lorentz force of the form%
\begin{equation}
m\ddot{x}^{\mu }\left( \tau \right) =e_{0}f^{\mu \nu }\left( x,\tau \right) 
\dot{x}_{\nu }\left( \tau \right) +e_{0}\varepsilon ^{\mu }\left( x,\tau
\right)  \label{Lorentz-Stk}
\end{equation}%
where $f^{\mu \nu }\left( x,\tau \right) $ is the electromagnetic field
tensor (made $\tau $-dependent), and $\varepsilon ^{\mu }\left( x,\tau
\right) $ is a field strength vector introduced to permit continuous
reversal of the time direction $\dot{x}^{0}\left( \tau \right) $. Without
the additional vector field,$\ $the value of $\dot{x}^{2}$ is conserved
through 
\begin{equation}
\frac{d}{d\tau }\left( \dot{x}^{2}\right) =2\dot{x}_{\mu }\ddot{x}^{\mu }=%
\frac{2e}{m}\dot{x}_{\mu }f^{\mu \nu }\dot{x}_{\nu }=0
\end{equation}%
and so the particle cannot evolve smoothly from $\dot{x}^{0}>0$ to $\dot{x}%
^{0}<0$ through spacelike motion with $\left( \dot{x}^{0}\right) ^{2}<%
\mathbf{\dot{x}}^{2}$. But Stueckelberg found no physical justification for
the vector field and proceeded to develop the $\tau $-parameterized
covariant canonical quantum mechanics%
\begin{equation}
i\partial _{\tau }\psi \left( x,\tau \right) =\frac{1}{2m}\left[ p^{\mu
}-eA^{\mu }\left( x\right) \right] \left[ p_{\mu }-eA_{\mu }\left( x\right) %
\right] \psi \left( x,\tau \right)  \label{K-Stk}
\end{equation}%
later used by Feynman \cite{Feynman}, Schwinger \cite{Schwinger}, DeWitt 
\cite{DeWitt} and others in developing QED. Stueckelberg required an
independent parameter because, by construction, Einstein coordinate time $%
x^{0}$ does not increase monotonically, and he regarded $\tau $ as a
physical time of system evolution, playing the role of Newtonian time in
nonrelativistic mechanics. The Hamiltonian system associated with (\ref%
{K-Stk}) is then a symplectic mechanics for which manifest Poincar\'{e}
covariance plays the role of Galilean covariance in Newtonian mechanics (see
also \cite{Fock,Nambu}).

Stueckelberg's generalized Lorentz force was later obtained from first
principles in the framework of a fundamental gauge theory \cite{H-P,saad},
called off-shell electrodynamics because the non-conservation of $\dot{x}%
^{2}\left( \tau \right) $ found from (\ref{Lorentz-Stk}) is associated with
mass exchange between particle and field. Quite generally \cite{beyond},
this framework is implicitly incorporated as the underlying electrodynamics
whenever one writes equations of motion%
\begin{equation}
m\ddot{x}^{\mu }=F^{\mu }\left( x,\dot{x},\tau \right)  \label{eq-motion}
\end{equation}%
on the unconstrained phase space defined by%
\begin{equation}
\left[ x^{\mu },x^{\nu }\right] =0\mbox{\qquad}\mbox{\qquad}m\left[ x^{\mu },%
\dot{x}^{\nu }\right] =-i\eta ^{\mu \nu },  \label{com-rel}
\end{equation}%
with metric 
\begin{equation}
\eta ^{\mu \nu }=\mathrm{diag}(-1,1,1,1).
\end{equation}%
and $\mu ,\nu =0,\cdots ,3$. The apparently naive commutation relations (\ref%
{com-rel}) are actually sufficient \cite{H-S} to establish the
self-adjointness of (\ref{eq-motion}) from which one is necessarily led to
the Lorentz force (\ref{K-Stk}) along with homogeneous field equations%
\begin{equation}
\partial _{\mu }f_{\nu \rho }+\partial _{\nu }f_{\rho \mu }+\partial _{\rho
}f_{\mu \nu }=0\mbox{\qquad}\partial _{\mu }\varepsilon _{\nu }-\partial
_{\nu }\varepsilon _{\mu }+\partial _{\tau }f_{\mu \nu }=0.  \label{sum3}
\end{equation}%
The fields $f_{\mu \nu }\left( x,\tau \right) $ and $\varepsilon _{\mu
}\left( x,\tau \right) $ have been called pre-Maxwell fields because of
their explicit $\tau $-dependence. Adopting formal designations 
\begin{equation}
x^{5}=\tau \mbox{\qquad}\partial _{5}=\partial _{\tau }\mbox{\qquad}%
\varepsilon _{\mu }=f_{5\mu }
\end{equation}%
and%
\begin{equation}
\mu ,\nu =0,\cdots ,3\mbox{\qquad}\mbox{\qquad}\alpha ,\beta =0,1,2,3,5
\end{equation}%
we summarize the Lorentz force (\ref{Lorentz-Stk}) and field equations (\ref%
{sum3}) as 
\begin{eqnarray}
m\ddot{x}_{\mu }=e_{0}f_{\mu \alpha }\left( x,\tau \right) \dot{x}^{\alpha }%
\mbox{\quad} &\Rightarrow &\mbox{\quad}\frac{d}{d\tau }(-\tfrac{1}{2}m\dot{x}%
^{2})=e_{0}f_{5\alpha }\left( x,\tau \right) \dot{x}^{\alpha }  \label{Lor}
\\
\rule[-0.3cm]{0cm}{1.2cm}\partial _{\lbrack \alpha }f_{\beta \gamma ]}=0%
\mbox{\quad} &\Rightarrow &\mbox{\quad}f_{\alpha \beta }\left( x,\tau
\right) =\partial _{\alpha }a_{\beta }\left( x,\tau \right) -\partial
_{\beta }a_{\alpha }\left( x,\tau \right)  \label{fld}
\end{eqnarray}%
from which the Lagrangian is uniquely determined \cite{Santilli} as%
\begin{equation}
L=\tfrac{1}{2}m\dot{x}^{\mu }\dot{x}_{\mu }+e_{0}\,\dot{x}^{\alpha
}\,a_{\alpha }\left( x,\tau \right) .  \label{act}
\end{equation}%
Writing the canonical momentum%
\begin{equation}
p_{\mu }=\frac{\partial L}{\partial \dot{x}^{\mu }}=m\dot{x}_{\mu
}+e_{0}a_{\mu }\left( x,\tau \right)
\end{equation}%
and transforming to the Hamiltonian%
\begin{equation}
K=p_{\mu }\dot{x}^{\mu }-L
\end{equation}%
the Stueckelberg-Schrodinger equation%
\begin{equation}
\left[ i\partial _{\tau }+e_{0}a_{5}\left( x,\tau \right) \right] \psi
\left( x,\tau \right) =\frac{1}{2m}\left[ p^{\mu }-e_{0}a^{\mu }\left(
x,\tau \right) \right] \left[ p_{\mu }-e_{0}a_{\mu }\left( x,\tau \right) %
\right] \psi \left( x,\tau \right)  \label{St-Sh}
\end{equation}%
is seen to be locally gauge invariant \cite{saad} under $\tau $-dependent
gauge transformations of the type 
\begin{equation}
\psi \rightarrow e^{ie_{0}\Lambda \left( x,\tau \right) }\psi \mbox{\qquad}%
a_{\mu }\rightarrow a_{\mu }+\partial _{\mu }\Lambda \left( x,\tau \right) %
\mbox{\qquad}a_{5}\rightarrow a_{5}+\partial _{\tau }\Lambda \left( x,\tau
\right) .
\end{equation}%
The electromagnetic field $f_{\alpha \beta }\left( x,\tau \right) $ is made
a dynamical quantity by adding a kinetic term to the action. In analogy to
standard Maxwell theory, one may adopt the formal designation $f^{\mu
5}=\eta ^{55}f_{~~5}^{\mu }=-f_{~~5}^{\mu }$ and chose the form%
\begin{equation}
-\frac{\lambda }{4}f^{\alpha \beta }\left( x,\tau \right) f_{\alpha \beta
}\left( x,\tau \right)
\end{equation}%
which is a gauge invariant Lorentz scalar and contains only first order
derivatives of the fields. The electromagnetic part of the action is now%
\begin{equation}
S_{field}=\int d^{4}xd\tau ~\left\{ e_{0}\dot{z}^{\alpha }a_{\alpha }\left(
x,\tau \right) \delta ^{4}\left[ x-z\left( \tau \right) \right] -\frac{%
\lambda }{4}f^{\alpha \beta }\left( x,\tau \right) f_{\alpha \beta }\left(
x,\tau \right) \right\}  \label{Action-1}
\end{equation}%
leading to inhomogeneous field equations 
\begin{equation}
\partial _{\beta }f^{\alpha \beta }\left( x,\tau \right) =\frac{e_{0}}{%
\lambda }j^{\alpha }\left( x,\tau \right) =ej^{\alpha }\left( x,\tau \right)
=e\dot{z}^{\alpha }\left( \tau \right) \delta ^{4}\left[ x-z\left( \tau
\right) \right]  \label{pM}
\end{equation}%
and a conserved 5-current 
\begin{equation}
\partial _{\mu }j^{\mu }\left( x,\tau \right) +\partial _{\tau }j^{5}\left(
x,\tau \right) =0.  \label{current}
\end{equation}%
As in nonrelativistic mechanics, equation (\ref{current}) suggests the
interpretation of $j^{5}\left( x,\tau \right) $ as the probability density
at $\tau $ of finding the event at $x$. Since $\partial _{\mu }j^{\mu }\neq
0 $, $j^{\mu }\left( x,\tau \right) $ cannot be identified as the source
current in Maxwell's equations. However, under the boundary conditions $%
j^{5}\rightarrow 0$, pointwise, as $\tau \rightarrow \pm \infty $,
integration of (\ref{current}) over $\tau $, leads to $\partial _{\mu
}J^{\mu }=0$, where 
\begin{equation}
J^{\mu }(x)=\int_{-\infty }^{\infty }d\tau \ j^{\mu }\left( x,\tau \right) \
.  \label{eqn:8}
\end{equation}%
This integration has been called concatenation \cite{concat} and links the
event current $j^{\mu }\left( x,\tau \right) $ with the particle current $%
J^{\mu }(x)$ defined on the entire particle worldline. Similarly, 
\begin{equation}
\left. 
\begin{array}{c}
\partial _{\beta }f^{\alpha \beta }\left( x,\tau \right) =ej^{\alpha }\left(
x,\tau \right) \\ 
\\ 
\partial _{\lbrack \alpha }f_{\beta \gamma ]}=0%
\end{array}%
\right\} \underset{\int d\tau }{\mbox{\quad}\xrightarrow{\hspace*{1cm}}%
\mbox{\quad}}\left\{ 
\begin{array}{c}
\partial _{\nu }F^{\mu \nu }\left( x\right) =eJ^{\mu }\left( x\right) \\ 
\\ 
\partial _{\lbrack \mu }F_{\nu \rho ]}=0%
\end{array}%
\right.
\end{equation}%
where 
\begin{equation}
F^{\mu \nu }(x)=\int_{-\infty }^{\infty }d\tau f^{\mu \nu }\left( x,\tau
\right) \mbox{\qquad}A^{\mu }(x)=\int_{-\infty }^{\infty }d\tau a^{\mu
}\left( x,\tau \right)  \label{c-cat-2}
\end{equation}%
extracting standard Maxwell theory as the equilibrium limit of event
dynamics. It is seen from (\ref{St-Sh}) and (\ref{c-cat-2}) that $e_{0}$ and 
$\lambda $ must have dimensions of time, so that the dimensionless ratio $%
e=e_{0}/\lambda $ can be identified as the Maxwell charge.

As in the non-relativistic case, the two-body action-at-a-distance potential
in the Horwitz-Piron theory \cite{H-P} may be understood as the
approximation $-e_{0}a_{5}\left(x,\tau\right)\longrightarrow V(x)$. Within
this framework, solutions have been found for the generalizations of the
standard central force problem, including potential scattering \cite%
{scattering} and bound states \cite{I,II}. Examination of radiative
transitions \cite{selrul}, in particular the Zeeman \cite{zeeman} and Stark
effects \cite{stark}, indicate that all five components of the gauge
potential are necessary for an adequate explanation of observed
phenomenology.

The wave equation derived from (\ref{pM}) is 
\begin{equation}
\partial _{\alpha }\partial ^{\alpha }a^{\beta }\left( x,\tau \right)
=\left( \partial _{\mu }\partial ^{\mu }-\;\partial _{\tau }^{2}\right)
a^{\beta }\left( x,\tau \right) =-ej^{\beta }\left( x,\tau \right)
\label{wave}
\end{equation}%
for which the Greens function \cite{green} found from 
\begin{equation}
\left( \partial _{\mu }\partial ^{\mu }-\;\partial _{\tau }^{2}\right)
G\left( x,\tau \right) =-\delta ^{4}\left( x,\tau \right)
\end{equation}%
is%
\begin{equation}
G\left( x,\tau \right) =-{\frac{1}{{2\pi }}}\delta (x^{2})\delta \left( \tau
\right) -{\frac{1}{{2\pi ^{2}}}\frac{\partial }{{\partial {x^{2}}}}}\ {\frac{%
{\theta (x^{2}-\tau ^{2})}}{\sqrt{{x^{2}-\tau ^{2}}}}}=D\left( x\right)
\delta (\tau )-G_{correlation}\left( x,\tau \right) .  \label{grn}
\end{equation}%
The first term has support on the lightcone at instantaneous $\tau $, and
recovers the standard Maxwell Greens function under concatenation. The
second term has spacelike support (${x^{2}>\tau ^{2}\geq 0}$) and vanishes
under concatenation, so it may contribute to correlations but not to Maxwell
potentials. Potentials obtained from (\ref{grn}) as%
\begin{equation}
a^{\beta }\left( x,\tau \right) =-e\int d^{4}x^{\prime }d\tau ^{\prime
}~G\left( x-x^{\prime },\tau -\tau ^{\prime }\right) \ j^{\beta }(x^{\prime
},\tau ^{\prime })
\end{equation}%
recover Maxwell potentials under concatenation%
\begin{eqnarray}
A^{\mu }(x) &=&\int d\tau ~\left\{ -e\int d^{4}x^{\prime }d\tau ^{\prime
}~G\left( x-x^{\prime },\tau -\tau ^{\prime }\right) \ j^{\beta }(x^{\prime
},\tau ^{\prime })\right\}  \notag \\
&=&-e\int d^{4}x^{\prime }~D\left( x-x^{\prime }\right) \int d\tau ^{\prime
}\ j^{\mu }(x^{\prime },\tau ^{\prime })  \notag \\
&=&-e\int d^{4}x^{\prime }~D\left( x-x^{\prime }\right) J^{\mu }(x^{\prime
}).
\end{eqnarray}%
To study the low energy Coulomb problem, the source is taken to be the
\textquotedblleft static\textquotedblright\ event 
\begin{equation}
z\left( \tau \right) =\left( \tau ,0,0,0\right)
\end{equation}%
from which one calculates%
\begin{eqnarray}
a^{0}\left( x,\tau \right) &=&-e\int d^{4}x^{\prime }d\tau ^{\prime }D\left(
x-x^{\prime }\right) \delta \left( \tau -\tau ^{\prime }\right) \delta
\left( x^{0\prime }-\tau ^{\prime }\right) \delta ^{3}\left( \mathbf{x}%
\right)  \notag \\
&=&-{\frac{e}{{2\pi }}}\ \delta \left[ \left( x-\tau \right) ^{2}\right]
\theta \left( x^{0}\right)  \notag \\
&=&-{\frac{e}{{4\pi }\left\vert \mathbf{x}\right\vert }}\ \delta \left(
x^{0}-\tau -\left\vert \mathbf{x}\right\vert \right)  \label{a-delt}
\end{eqnarray}%
Although concatenation of (\ref{a-delt}) recovers the correct Coulomb
potential 
\begin{equation}
A^{0}(x)=-{\frac{e}{{4\pi }\left\vert \mathbf{x}\right\vert }}\int d\tau
~\delta \left( x^{0}-\tau -\left\vert \mathbf{x}\right\vert \right) =-{\frac{%
e}{{4\pi }\left\vert \mathbf{x}\right\vert }}
\end{equation}%
the $\delta $-function in $a^{0}\left( x,\tau \right) $ results in a
microscopic dynamics that cannot reproduce expected low-energy interactions
between a pair of charges. In particular, a second \textquotedblleft
static\textquotedblright\ event given by 
\begin{equation}
\zeta \left( \tau \right) =\left( u^{0}\tau +\alpha ,\beta ,0,0\right)
\end{equation}%
will generally not experience any interaction with $z\left( \tau \right) $,
except for appropriately tuned values of $\dot{\zeta}^{0}=u^{0}$ and the
offset $\zeta ^{0}\left( 0\right) =\alpha $ that determines the $\tau $%
-synchronization of the two events.

Since off-shell quantum theory does provide a correct description of Coulomb
scattering for sharp asymptotic mass states, which contain no information
about $\tau $-synchronization, it was suggested in \cite{larry} that the
electromagnetic interaction between events should be modified to relax the
deterministic synchronization expressed in (\ref{Lor}) and (\ref{pM}). Under
this modification, the source $j_{\varphi }^{\beta }\left( x,\tau \right) $
for the pre-Maxwell fields is taken to be a smoothed current density induced
by an ensemble of events $z^{\beta }\left( \tau +\delta \tau \right) $ along
a particle worldline, where $\delta \tau $ is given by a normalized
distribution $\varphi \left( \tau \right) $. From (\ref{pM}) the smoothed
source current is given as the ensemble average%
\begin{equation}
j_{\varphi }^{\alpha }\left( x,\tau \right) =\int_{-\infty }^{\infty }ds\
\varphi \left( \tau -s\right) \dot{z}^{\alpha }\left( s\right) \delta ^{4}%
\left[ x-z\left( s\right) \right]
\end{equation}%
or equivalently as an ensemble of sharp currents $j^{\beta }\left( x,\tau
\right) $ induced by a single event%
\begin{equation}
j^{\alpha }\left( x,\tau \right) \longrightarrow j_{\varphi }^{\alpha
}\left( x,\tau \right) =\int_{-\infty }^{\infty }ds\ \varphi \left( \tau
-s\right) ~j^{\alpha }\left( x,s\right) .  \label{fi-cur}
\end{equation}%
The fields $f^{\alpha \beta }\left( x,\tau \right) $ and $a^{\beta }\left(
x,\tau \right) $ are induced by the smoothed current density, so the
electrodynamic system that mediates between sharp events $z^{\beta }\left(
\tau \right) $ and $\zeta ^{\beta }\left( \tau \right) $ explicitly
introduces a statistical structure to the event-event interaction. Because
Maxwell currents are defined along the entire worldline, this procedure
preserves the concatenated current 
\begin{equation}
J^{\mu }\left( x\right) =\int_{-\infty }^{\infty }d\tau \ j_{\varphi }^{\mu
}\left( x,\tau \right) =\int_{-\infty }^{\infty }ds\left[ \int_{-\infty
}^{\infty }d\tau \varphi \left( \tau -s\right) j^{\mu }\left( x,s\right) %
\right] =\int_{-\infty }^{\infty }ds\ j^{\mu }\left( x,s\right) .
\end{equation}%
Taking the distribution to be%
\begin{equation}
\varphi \left( \tau \right) =\frac{1}{2\lambda }e^{-|\tau |/\lambda }%
\mbox{\qquad}\mbox{\qquad}\int_{-\infty }^{\infty }d\tau ~\varphi \left(
\tau \right) =1  \label{fi-def}
\end{equation}%
the low energy Coulomb field becomes a Yukawa-type potential with the
correct non-relativistic limit for large $\lambda $%
\begin{equation}
m\mathbf{\ddot{x}}=-e_{0}\nabla \left[ a^{0}\left( x,\tau \right)
+a^{5}\left( x,\tau \right) \right] \rightarrow m\mathbf{\ddot{x}}%
=-2e_{0}\nabla a_{\varphi }^{0}\left( x,\tau \right) =e^{2}\nabla \left[ 
\frac{e^{-\left\vert \mathbf{x}\right\vert /\lambda }}{{4\pi }\left\vert 
\mathbf{x}\right\vert }\right] .
\end{equation}%
The distribution $\varphi \left( \tau \right) $ provides a cutoff for the
photon mass spectrum, which we take to be the conventional experimental
uncertainty in photon mass ($\Delta m_{\gamma }\simeq 10^{-17}\unit{eV}$ 
\cite{pdg}), leading to a value of about 400 seconds for $\lambda $. The
limit $\lambda \rightarrow 0$ restores $\varphi \left( \tau \right)
\rightarrow \delta \left( \tau \right) $ and the limit $\lambda \rightarrow
\infty $ restores standard Maxwell theory. Since the form of $\varphi \left(
\tau \right) $ given in (\ref{fi-cur}) represents the distribution of
interarrival times of events in a Poisson-distributed stochastic process,
this choice suggests an information-theoretic interpretation for the
underlying the relationship between the current density and the ensemble of
events from which it is induced.

The smoothed current can be introduced through the action \cite{high-order},
by adding a higher $\tau $-derivative term to the electromagnetic part. The
substitution 
\begin{equation}
S_{em}\rightarrow \int d^{4}xd\tau \left[ e_{0}j^{\alpha }a_{\alpha }-\frac{%
\lambda }{4}f^{\alpha \beta }\left( x,\tau \right) f_{\alpha \beta }\left(
x,\tau \right) -\frac{\lambda ^{3}}{4}\left[ \partial _{\tau }f^{\alpha
\beta }\left( x,\tau \right) \right] \left[ \partial _{\tau }f_{\alpha \beta
}\left( x,\tau \right) \right] \right]
\end{equation}%
preserves Lorentz and gauge invariance, and leaves the action first order in
spacetime derivatives. Defining a field interaction kernel 
\begin{equation}
\Phi \left( \tau \right) =\delta \left( \tau \right) -\lambda ^{2}\delta
^{\prime \prime }\left( \tau \right) =\frac{1}{2\pi }\int d\kappa \,\left[
1+\left( \lambda \kappa \right) ^{2}\right] \,e^{-i\kappa \tau }
\end{equation}%
which is seen from%
\begin{equation}
\int_{-\infty }^{\infty }ds~\Phi (\tau -s)\varphi (s)=\delta (\tau
)\rightarrow \varphi \left( \tau \right) =\int \frac{d\kappa }{2\pi }\frac{%
e^{-i\kappa \tau }}{1+\left( \lambda \kappa \right) ^{2}}=\frac{1}{2\lambda }%
e^{-|\tau |/\lambda }  \label{inverse}
\end{equation}%
to be the inverse function to $\varphi (\tau )$, the action becomes%
\begin{equation}
S_{em}=\int d^{4}xd\tau ~e_{0}j^{\alpha }a_{\alpha }-\frac{\lambda }{4}\int
d^{4}x\,d\tau \,ds\ f^{\alpha \beta }\left( x,\tau \right) \Phi (\tau
-s)f_{\alpha \beta }\left( x,s\right) .  \label{Action-2}
\end{equation}%
The Euler-Lagrange equations 
\begin{equation}
\partial _{\beta }f_{\Phi }^{\alpha \beta }(x,\tau )=\partial _{\beta }\int
ds\,\Phi (\tau -s)f^{\alpha \beta }(x,s)=ej^{\alpha }\left( x,\tau \right)
\label{EuM}
\end{equation}%
describe a sharp field induced by a sharp event current, and using (\ref%
{inverse}) can be inverted to recover%
\begin{equation}
\partial _{\beta }f^{\alpha \beta }\left( x,\tau \right) =ej_{\varphi
}^{\alpha }\left( x,\tau \right) =e\int ds~\varphi \left( \tau -s\right)
j^{\alpha }\left( x,s\right) .  \label{pMm}
\end{equation}%
The action (\ref{Action-2}), in which the statistical synchronization
performed by $\Phi (\tau -s)$ is made explicit, has the advantage of
permitting the usual study of symmetries and being amenable to second
quantization, where the factor $\left[ 1+\left( \lambda \kappa \right) ^{2}%
\right] ^{-1}$ provides a natural mass cutoff for the off-shell photon that
renders off-shell quantum field theory super-renormalizable at two-loop
order \cite{high-order}.

\section{Radiation Fields}

In this section we calculate the electromagnetic field induced by an
arbitrarily evolving spacetime event, in order to identify and characterize
the radiation part. In particular, we obtain the five Li\'{e}nard-Wiechert
potentials and the field strength tensor, which is conveniently expressed as
a Clifford product of a pair orthogonal vectors. This bivector formulation
simplifies calculation of field invariants, the mass-energy-momentum tensor,
and the plane wave expansion. The radiation field is shown to be an
equilibrium field of the standard Maxwell type.

\subsection{Li\'{e}nard-Wiechert potential}

Beginning with a generic spacetime event $X^{\mu }\left( \tau \right) $ and
using (\ref{pM}) and (\ref{fi-cur}), we write the smoothed current%
\begin{equation}
j_{\varphi }^{\alpha }\left( x,\tau \right) =\int ds~\varphi \left( \tau
-s\right) j^{\alpha }\left( x,s\right) =\int ds~\varphi \left( \tau
-s\right) \dot{X}^{\alpha }\left( s\right) \delta ^{4}\left[ x-X\left(
s\right) \right]  \label{cur-1}
\end{equation}%
where%
\begin{equation}
\dot{X}^{5}\left( \tau \right) =\dot{\tau}=1.
\end{equation}%
We choose the $\tau $-instantaneous part of the Greens function (\ref{grn}),
which recovers the Maxwell propagator under concatenation (radiation
reaction associated with the correlation Greens function has been studied in 
\cite{Jigal} and references therein). The Li\'{e}nard-Wiechert potential $%
a^{\alpha }\left( x,\tau \right) $ induced by the current (\ref{cur-1}) is%
\begin{eqnarray}
a^{\alpha }\left( x,\tau \right) &&\mbox{\hspace{-20pt}}=-e\int
d^{4}x^{\prime }d\tau ^{\prime }G\left( x-x^{\prime },\tau -\tau ^{\prime
}\right) j_{\varphi }^{\alpha }\left( x^{\prime },\tau ^{\prime }\right) \\
&&\mbox{\hspace{-20pt}}=-\frac{e}{2\pi }\int ds~\varphi \left( \tau
-s\right) \dot{X}^{\alpha }\left( s\right) \delta \left[ \left( x-X\left(
s\right) \mbox{\rule{0cm}{10.5pt}}\right) ^{2}\right] \theta ^{ret}
\label{a_usefl} \\
&&\mbox{\hspace{-20pt}}=-\frac{e}{2\pi }\varphi \left( \tau -s\right) \frac{%
\dot{X}^{\alpha }\left( s\right) }{2\left( x^{\mu }-X^{\mu }\left( s\right)
\right) \dot{X}_{\mu }\left( s\right) }
\end{eqnarray}%
where we used the identity 
\begin{equation}
\dint d\tau f\left( \tau \right) \delta \left[ g\left( \tau \right) \right]
=\left. \dfrac{f\left( s\right) }{\left\vert g^{\prime }\left( s\right)
\right\vert }\right\vert _{s=g^{-1}\left( 0\right) }  \label{del-deriv}
\end{equation}%
and the retarded time $s$ satisfies 
\begin{equation}
\left[ x-X\left( s\right) \right] ^{2}=0\mbox{\qquad}\theta ^{ret}=\theta
\left( x^{0}-X^{0}\left( s\right) \right) =1.  \label{z-sq}
\end{equation}%
Introducing the timelike velocity%
\begin{equation}
u^{\alpha }=\dot{X}^{\alpha }\left( s\right) ,  \label{u}
\end{equation}%
the vector from event $X\left( s\right) $ to observation point $x$ 
\begin{equation}
z^{\mu }=x^{\mu }-X^{\mu }\left( s\right) \Rightarrow \dot{z}^{\mu }=-u^{\mu
},  \label{z-dot}
\end{equation}%
and the scalar function%
\begin{equation}
R=\frac{1}{2}\dfrac{d}{ds}\left( x-X\left( s\right) \right) ^{2}=-z^{\mu
}u_{\mu }=-z\cdot u\geq 0,  \label{R}
\end{equation}%
we find%
\begin{equation}
a^{\alpha }\left( x,\tau \right) =-e\frac{1}{2\lambda }\frac{u^{\alpha }}{%
4\pi R}e^{-|\tau -s|/\lambda }  \label{a_fnl}
\end{equation}%
where the nonnegativity of $R$ follows from (\ref{z-sq}) and (\ref{u}).

To calculate the field strengths, we need derivatives of the Li\'{e}%
nard-Wiechert potential. The spacetime derivative is most conveniently found
by applying the identity (\ref{del-deriv}) to expression (\ref{a_usefl})%
\begin{eqnarray}
\partial ^{\mu }a^{\beta }\left( x,\tau \right) &&\mbox{\hspace{-20pt}}=-%
\frac{e}{2\pi }\int ds~\varphi \left( \tau -s \right) \dot{X}^{\alpha
}\left( s\right) \theta ^{ret}\partial ^{\mu }\delta \left( \left( x-X\left(
s\right) \right) ^{2}\right) \\
&&\mbox{\hspace{-20pt}}={\frac{e}{{2\pi }}}\int ds\ \varphi (\tau -s )\dot{X}%
^{\beta }\left( s\right) \theta ^{ret}\delta ^{\prime }\left[ \left(
x-X\left( s\right) \right) ^{2}\right] \left[ -2\left( x^{\mu }-X^{\mu
}\left( s\right) \right) \right] \\
&&\mbox{\hspace{-20pt}}={\frac{e}{{2\pi }}}\int ds\ \varphi (\tau -s )\frac{%
\dot{X}^{\beta }\left( s\right) \left[ x^{\mu }-X^{\mu }\left( s\right) %
\right] }{\dot{X}\left( s\right) \cdot \left( x-X\left( s\right) \right) }%
\theta ^{ret}\frac{d}{ds}\delta \left[ \left( x-X\left( s\right) \right) ^{2}%
\right]
\end{eqnarray}%
and integrating by parts to obtain%
\begin{eqnarray}
\partial ^{\mu }a^{\beta }\left( x,\tau \right) &&\mbox{\hspace{-20pt}}=-{%
\frac{e}{{2\pi }}}\int ds\frac{d}{ds}\left[ \varphi (\tau -s )\frac{\dot{X}%
^{\beta }\left( s\right) \left[ x^{\mu }-X^{\mu }\left( s\right) \right] }{%
\dot{X}\left( s\right) \cdot \left( x-X\left( s\right) \right) }\right]
\theta ^{ret}\delta \left[ \left( x-X\left( s\right) \right) ^{2}\right]
\label{der-0} \\
&&\mbox{\hspace{-20pt}}=-{\frac{e}{{4\pi }}}\frac{1}{R}\frac{d}{ds}\left[
\varphi (\tau -s )\frac{z^{\mu }u^{\beta }}{R}\right] .  \label{der-1}
\end{eqnarray}%
Since%
\begin{equation}
\frac{d}{d\tau }\varphi \left( \tau \right) =\frac{1}{2\lambda }\frac{d}{%
d\tau }e^{-\left\vert \tau \right\vert /\lambda }=-\frac{1}{2\lambda ^{2}}%
\epsilon \left( \tau \right) e^{-\left\vert \tau \right\vert /\lambda }=-%
\frac{1}{\lambda }\epsilon \left( \tau \right) \varphi \left( \tau \right)
\label{d-fi}
\end{equation}%
we obtain the $\tau $-derivative directly from (\ref{a_fnl}) as 
\begin{equation}
\partial _{\tau }a_{\mu }\left( x,\tau \right) =-e\dot{\varphi}\left( \tau
-s \right) \frac{u_{\mu }}{4\pi R}=e\epsilon \left( \tau -s\right) \varphi
\left( \tau -s \right) \frac{u_{\mu }}{4\pi \lambda R}~~.  \label{der-2}
\end{equation}

\subsection{Field strengths}

From (\ref{der-1}) the spacetime components of the field strength tensor
takes the anti-symmetric form%
\begin{equation}
f^{\mu \nu }=-{\frac{e}{{4\pi }}}\frac{1}{R}\frac{d}{ds}\left[ \varphi (\tau
-s )\frac{z^{\mu }u^{\nu }-z^{\nu }u^{\mu }}{R}\right] .  \label{f-1}
\end{equation}%
Using (\ref{z-dot}), (\ref{R}), and%
\begin{equation}
\dot{R}=-\frac{d}{d\tau }\left( z\cdot u\right) =u^{2}-z\cdot \dot{u}
\label{R-dot}
\end{equation}%
the derivatives split into a retarded field exhibiting $\left\vert \mathbf{z}%
\right\vert ^{-2}$ far-field behavior%
\begin{equation}
f_{ret}^{\mu \nu }=e\varphi \left( \tau -s \right) \left[ \frac{\left(
z^{\mu }u^{\nu }-z^{\nu }u^{\mu }\right) u^{2}}{{4\pi }\left( u\cdot
z\right) ^{3}}-\epsilon \left( \tau -s\right) \frac{\left( z^{\mu }u^{\nu
}-z^{\nu }u^{\mu }\right) }{{4\pi }\lambda \left( u\cdot z\right) ^{2}}%
\right] ~,
\end{equation}%
and a long-range radiation field exhibiting $\left\vert \mathbf{z}%
\right\vert ^{-1}$ far-field behavior 
\begin{equation}
f_{rad}^{\mu \nu }=e\varphi (\tau -s )\frac{\left( z^{\mu }\dot{u}^{\nu
}-z^{\nu }\dot{u}^{\mu }\right) \left( u\cdot z\right) -\left( z^{\mu
}u^{\nu }-z^{\nu }u^{\mu }\right) \left( \dot{u}\cdot z\right) }{{4\pi }%
\left( u\cdot z\right) ^{3}}~.  \label{rad-1}
\end{equation}%
Because we take $\lambda \simeq $ 400 seconds, we include the $\lambda ^{-1}$
term in the retarded field. Using (\ref{der-1}) to find $\partial _{\mu
}a_{5}$ and (\ref{der-2}) to find $\partial _{\tau }a_{\mu }$ we calculate
the fifth component fields as%
\begin{eqnarray}
f_{\mu 5}^{ret} &&\mbox{\hspace{-20pt}}=-e\varphi \left( \tau -s \right) %
\left[ \frac{z_{\mu }u^{2}-u_{\mu }\left( u\cdot z\right) }{{4\pi }\left(
u\cdot z\right) ^{3}}-\epsilon \left( \tau -s\right) \frac{z_{\mu }-u_{\mu
}\left( u\cdot z\right) }{{4\pi }\lambda \left( u\cdot z\right) ^{2}}\right]
\\
f_{\mu 5}^{rad} &&\mbox{\hspace{-20pt}}=-e\varphi \left( \tau -s \right) 
\frac{\left( \dot{u}\cdot z\right) z_{\mu }}{{4\pi }\left( u\cdot z\right)
^{3}}~.  \label{rad-2}
\end{eqnarray}

\subsection{Electromagnetic two form}

For notational simplicity, we will represent the field as a bivector%
\begin{equation}
f=\frac{1}{2}f^{\alpha \beta }\left( \mathbf{e}_{\alpha }\wedge \mathbf{e}%
_{\beta }\right) =\frac{1}{2}f^{\alpha \beta }\left( \mathbf{e}_{\alpha
}\otimes \mathbf{e}_{\beta }-\mathbf{e}_{\beta }\otimes \mathbf{e}_{\alpha
}\right)
\end{equation}%
in a Clifford algebra (see \cite{cliff} and references contained therein)
over the formal 5D space with basis vectors%
\begin{equation}
\mathbf{e}_{\alpha }\cdot \mathbf{e}_{\beta }=\eta _{\alpha \beta }%
\mbox{\qquad}\alpha ,\beta =0,1,2,3,5.
\end{equation}%
In this notation, expression (\ref{f-1}) for the spacetime component of the
electromagnetic field becomes%
\begin{equation}
f_{spacetime}=\frac{1}{2}f^{\mu \nu }\left( \mathbf{e}_{\mu }\wedge \mathbf{e%
}_{\nu }\right) =-{\frac{e}{{4\pi }}}\frac{1}{R}\frac{d}{ds}\left[ \varphi
(\tau -s)\frac{z\wedge u}{R}\right]
\end{equation}%
and the radiation part (\ref{rad-1}) is%
\begin{eqnarray}
f_{spacetime} &&\mbox{\hspace{-20pt}}=e\varphi (\tau -s)\frac{\left( z\wedge 
\dot{u}\right) \left( u\cdot z\right) -\left( z\wedge u\right) \left( \dot{u}%
\cdot z\right) }{{4\pi }\left( u\cdot z\right) ^{3}} \\
&&\mbox{\hspace{-20pt}}=e\varphi (\tau -s)~z\wedge \frac{\dot{u}\left(
u\cdot z\right) -u\left( \dot{u}\cdot z\right) }{{4\pi }\left( u\cdot
z\right) ^{3}}~.
\end{eqnarray}%
Similarly, the fifth component radiation field (\ref{rad-2}) takes the form%
\begin{equation}
f_{5}=\frac{1}{2}f^{\mu 5}\left( \mathbf{e}_{\mu }\wedge \mathbf{e}%
_{5}\right) =-e\varphi \left( \tau -s\right) \frac{\left( \dot{u}\cdot
z\right) ~z\wedge \mathbf{e}_{5}}{{4\pi }\left( u\cdot z\right) ^{3}}~.
\end{equation}%
Introducing the scalar quantity%
\begin{equation}
Q=-\dot{u}\cdot z  \label{Q}
\end{equation}%
and the vector%
\begin{equation}
w=\dot{u}R-uQ=-\left[ \dot{u}\left( u\cdot z\right) -u\left( \dot{u}\cdot
z\right) \right]  \label{w}
\end{equation}%
the radiation fields assume the form%
\begin{eqnarray}
f_{spacetime} &&\mbox{\hspace{-20pt}}=e\varphi (\tau -s)~\frac{z\wedge w}{{%
4\pi }R^{3}} \\
f_{5}^{rad} &&\mbox{\hspace{-20pt}}=-e\varphi \left( \tau -s\right) ~\frac{%
z\wedge \left( Q\mathbf{e}_{5}\right) }{{4\pi }R^{3}}~.
\end{eqnarray}%
Designating formal 5-velocity and 5-acceleration as%
\begin{equation}
U=\frac{d}{d\tau }\left( X\left( \tau \right) ,\tau \right) =\left( \dot{X}%
\left( \tau \right) ,1\right) =u+\mathbf{e}_{5}\mbox{\qquad}\dot{U}=\frac{d}{%
d\tau }\left( \dot{X}\left( \tau \right) ,1\right) =\left( \dot{u},0\right) =%
\dot{u}  \label{U}
\end{equation}%
so that%
\begin{equation}
R=-z\cdot U\mbox{\qquad}Q=-z\cdot \dot{U}  \label{RQ}
\end{equation}%
the radiation fields may be combined in the expression 
\begin{equation}
f=f_{spacetime}+f_{5}=e\varphi (\tau -s)~\frac{z\wedge W}{{4\pi }R^{3}}.
\label{fr-2}
\end{equation}%
where the 5D vector%
\begin{equation}
W=\dot{U}R-UQ=w-Q\mathbf{e}_{5}  \label{W}
\end{equation}%
depends on the velocity and acceleration of the event, characterizing the
directionality of the motion inducing the radiation.

\subsection{Field invariants}

From (\ref{RQ}) and the Clifford identity%
\begin{equation}
a\cdot \left( b\wedge c\right) =\left( a\cdot b\right) c-\left( a\cdot
c\right) b  \label{Cl-1}
\end{equation}%
the vector $W$ can be expressed as%
\begin{equation}
W=U\left( z\cdot \dot{U}\right) -\dot{U}\left( z\cdot U\right) =z\cdot
\left( U\wedge \dot{U}\right)
\end{equation}%
describing the projection of $z$ into the plane spanned by the vectors $U$
and $\dot{U}$, and reflected through the direction of $U$. The vector $W$ is
orthogonal to $z$, which follows from the Clifford identity%
\begin{equation}
\left( a\wedge b\right) \cdot \left( c\wedge d\right) =\left[ \left( a\wedge
b\right) \cdot c\right] \cdot d=\left( a\cdot c\right) \left( b\cdot
d\right) -\left( b\cdot c\right) \left( a\cdot d\right)  \label{Cl-2}
\end{equation}%
as%
\begin{equation}
z\cdot W=z\cdot \left[ z\cdot \left( U\wedge \dot{U}\right) \right] =\left(
z\wedge z\right) \cdot \left( U\wedge \dot{U}\right)
\end{equation}%
or directly as 
\begin{equation}
z\cdot W=z\cdot \left( \dot{U}R-UQ\right) =\left( z\cdot \dot{U}\right)
R-\left( z\cdot U\right) Q=-QR+RQ=0.  \label{z.W}
\end{equation}%
The radiation field thus lies in the plane spanned by the orthogonal vectors 
$z$ and $W$, where $W$ is normal to $z$ in the plane spanned by $U$ and $%
\dot{U}$. It also follows from (\ref{z.W}), (\ref{z-sq}) and the Clifford
identity (\ref{Cl-1}) that%
\begin{equation}
z\cdot \left( z\wedge W\right) =z^{2}W-z\left( z\cdot W\right) =0
\label{z.W-2}
\end{equation}%
so that the plane $z\wedge W$ is null with respect to the lightlike
observation vector $z$. Writing the components of the field bivector (\ref%
{fr-2}) using%
\begin{equation}
\left( z\wedge W\right) _{\beta \gamma }=z_{\beta }W_{\gamma }-z_{\gamma
}W_{\beta }
\end{equation}%
the field invariant $\epsilon ^{\alpha \beta \gamma \delta \varepsilon
}f_{\beta \gamma }f_{\delta \varepsilon }$ is seen to vanish identically
because the product of vectors satisfies%
\begin{equation}
\epsilon ^{\alpha \beta \gamma \delta \varepsilon }\left( z_{\beta
}W_{\gamma }-z_{\gamma }W_{\beta }\right) \left( z_{\delta }W_{\varepsilon
}-z_{\varepsilon }W_{\delta }\right) =4\epsilon ^{\alpha \beta \gamma \delta
\varepsilon }z_{\beta }W_{\gamma }z_{\delta }W_{\varepsilon }=0.
\end{equation}%
Similarly, since $f^{\alpha \beta }f_{\alpha \beta }$ includes $\left(
z\wedge W\right) ^{\alpha \beta }\left( z\wedge W\right) _{\alpha \beta }$
it must vanish because using (\ref{Cl-2}) we find%
\begin{equation}
\left( z\wedge W\right) \cdot \left( z\wedge W\right) =\left( z\cdot
W\right) ^{2}-z^{2}W^{2}=0.
\end{equation}%
Thus, the radiation field (\ref{fr-2}) is seen to be a null field, satisfying%
\begin{eqnarray}
z\cdot f &&\mbox{\hspace{-20pt}}=0\;\longrightarrow \;z_{\mu }f^{\mu \alpha
}~=0  \label{inv-1} \\
f\cdot f &&\mbox{\hspace{-20pt}}=0\;\longrightarrow \;f^{\alpha \beta
}f_{\alpha \beta }=f^{\mu \nu }f_{\mu \nu }=f^{5\nu }f_{5\nu }=0%
\begin{array}{c}
~ \\ 
~%
\end{array}
\label{inv-2} \\
\epsilon ^{\alpha \beta \gamma \delta \varepsilon }f_{\beta \gamma
}f_{\delta \varepsilon } &&\mbox{\hspace{-20pt}}=0\;\longrightarrow
\;\epsilon ^{\mu \nu \rho \sigma }f_{\mu \nu }f_{\rho \sigma }=\epsilon
^{\mu \nu \rho \sigma }f_{\sigma 5}f_{\rho \nu }=0.  \label{inv-3}
\end{eqnarray}

\subsection{Mass-energy-momentum tensor}

In \cite{saad} the mass-energy-momentum tensor was derived as the 
\QTR{framesubtitle}{Noether current associated with the }translation
invariance of electromagnetic action in the form (\ref{Action-1}). Studying
the translation invariance of the modified action (\ref{Action-2}), one
finds the conservation law%
\begin{equation}
\partial _{\alpha }T^{\alpha \beta }=e_{0}f^{\alpha \beta }j_{\alpha }
\label{dT}
\end{equation}%
for conserved tensor%
\begin{equation}
T^{\alpha \beta }=-\lambda \left( g^{\alpha \beta }f^{\delta \gamma
}f_{\delta \gamma }^{\Phi }-f_{~\gamma }^{\alpha }f_{\Phi }^{\beta \gamma
}\right) ,  \label{T-ab}
\end{equation}%
where the modified field $f_{\Phi }$ is given in (\ref{EuM}) and the sharp
current $j_{\alpha }$ is defined in (\ref{pM}). Integrating (\ref{dT}) over
spacetime, leads to%
\begin{equation}
\frac{d}{d\tau }\int d^{4}x~T^{5\alpha }=e_{0}\int d^{4}x~f^{\alpha \beta
}\left( x,\tau \right) \dot{X}_{\alpha }\delta ^{4}\left( x-X\right)
=e_{0}f^{\alpha \beta }\left( X,\tau \right) \dot{X}_{\alpha }
\end{equation}%
which by comparison with the Lorentz force (\ref{Lor}) provides%
\begin{equation}
\frac{d}{d\tau }\int d^{4}x~\left( T^{5\mu }+m\dot{X}^{\mu }\right) =0%
\mbox{\qquad}\mbox{\qquad}\frac{d}{d\tau }\int d^{4}x\left( T^{55}-\tfrac{1}{%
2}m\dot{X}^{2}\right) =0,
\end{equation}%
expressing the instantaneous conservation of total energy-momentum-mass for
the combined field and event over all spacetime \cite{emlf}. Since the $\tau 
$-dependence of (\ref{fr-2}) resides in $\varphi \left( \tau -s\right) $,
the $\tau $-integration in (\ref{EuM}) merely inverts $\varphi \left( \tau
-s\right) $ to $\delta \left( \tau -s\right) $, and since%
\begin{equation}
\delta \left( \tau -s\right) \varphi \left( \tau -s\right) =\delta \left(
\tau -s\right) \varphi \left( 0\right) =\frac{1}{2\lambda }\delta \left(
\tau -s\right)
\end{equation}%
terms of the type $f^{\alpha \beta }f_{\delta \gamma }^{\Phi }$ become%
\begin{equation}
f^{\alpha \beta }f_{\delta \gamma }^{\Phi }=\frac{1}{2\lambda }\delta \left(
\tau -s\right) \left( \frac{e}{{4\pi }R^{3}}\right) ^{2}\left( z\wedge
W\right) ^{\alpha \beta }\left( z\wedge W\right) _{\delta \gamma }.
\end{equation}%
In light of (\ref{inv-2}), tensor (\ref{T-ab}) reduces to%
\begin{equation}
T^{\alpha \beta }=\lambda f_{~\gamma }^{\alpha }~f_{\Phi }^{\beta \gamma }
\end{equation}%
containing the products%
\begin{eqnarray}
\left( z^{\alpha }W_{\gamma }-z_{\gamma }W^{\alpha }\right) \left( z^{\beta
}W^{\gamma }-z^{\gamma }W^{\beta }\right) &&\mbox{\hspace{-20pt}}%
=z^{2}W^{\alpha }W^{\beta }+W^{2}z^{\alpha }z^{\beta }-z\cdot W\left(
z^{\alpha }W^{\beta }+z^{\alpha }W^{\beta }\right)  \notag \\
&&\mbox{\hspace{-20pt}}=W^{2}z^{\alpha }z^{\beta }
\end{eqnarray}%
and providing the simple expression%
\begin{equation}
T^{\alpha \beta }=\frac{1}{2}\delta (\tau -s)\left( \frac{e}{{4\pi }R^{3}}%
\right) ^{2}W^{2}z^{\alpha }z^{\beta }.
\end{equation}%
Because the observation vector $z$ has no 5-component, this becomes 
\begin{equation}
T^{\alpha 5}=0\mbox{\qquad}\mbox{\qquad}T^{\mu \nu }=\frac{1}{2}\delta (\tau
-s)\left( \frac{e}{{4\pi }R^{3}}\right) ^{2}W^{2}z^{\mu }z^{\nu }
\label{T-rad}
\end{equation}%
so that the total (over all spacetime) mass and energy-momentum~carried by
the radiation field vanishes.

Integrating (\ref{dT}) over space and $\tau $ --- equivalent to the space
integration of the concatenated dynamics --- we recover an expression for
the Maxwell Poynting vector 
\begin{equation}
\frac{d}{dt}\int d^{3}x~d\tau ~T^{0\mu }=e_{0}\int d^{3}x~d\tau ~f^{\alpha
\mu }\left( x,\tau \right) \dot{X}_{\alpha }\left( \tau \right) \delta ^{4}%
\left[ x-X\left( \tau \right) \right]  \label{PV}
\end{equation}%
where using (\ref{T-rad}), the energy-momentum%
\begin{equation}
P^{\mu }=\int d^{3}x~d\tau ~T^{0\mu }=\frac{1}{2}\left( \frac{e}{{4\pi }R^{3}%
}\right) ^{2}W^{2}z^{0}z^{\mu }
\end{equation}%
is found to be oriented along the observation vector $z^{\mu }$. We notice
that the RHS of (\ref{PV}) differs from the Maxwell formulation through the $%
\tau $-dependence of $f^{\alpha \beta }\left( x,\tau \right) $. In the
equilibrium limit $f^{\mu \nu }\left( x,\tau \right) \rightarrow \frac{1}{%
\lambda }F^{\mu \nu }\left( x\right) $, (\ref{PV}) is seen to recover the
standard form of energy-momentum~conservation 
\begin{equation}
\frac{d}{dt}\int d\tau ~T^{0\mu }=eF^{\mu \nu }\left( x\right) \int d\tau ~%
\dot{X}_{\nu }\left( \tau \right) \delta ^{4}\left[ x-X\left( \tau \right) %
\right] =eF^{\nu \mu }\left( x\right) J_{\nu }\left( x\right) .
\end{equation}

\subsection{Vector field picture}

To compare the radiation field (\ref{fr-2}) with the standard Maxwell field,
we write $f^{\alpha \beta }$ in vector components \cite{saad} as%
\begin{equation}
e^{i}=f^{0i}\mbox{\qquad}b^{i}=\epsilon ^{ijk}f_{jk}\mbox{\qquad}\varepsilon
^{\mu }=f^{5\mu }=\left( \varepsilon ^{0},\mathbf{\varepsilon }\right)
\end{equation}%
for which the field equations are%
\begin{equation}
\begin{array}{c}
\begin{array}{ccc}
\nabla \cdot \mathbf{e}-\partial _{\tau }\varepsilon ^{0}=ej^{0} & %
\mbox{\qquad\qquad\rule[-0.4cm]{0cm}{1cm}} & \nabla \times \mathbf{e}%
+\partial _{0}\mathbf{b}=0 \\ 
\nabla \times \mathbf{b}-\partial _{0}\mathbf{e}-\partial _{\tau }\mathbf{%
\varepsilon }=e\mathbf{j} & \mbox{\qquad\qquad\rule[-0.4cm]{0cm}{1cm}} & 
\nabla \cdot \mathbf{b}=0 \\ 
\nabla \cdot \mathbf{\varepsilon +}\partial _{0}\varepsilon ^{0}=ej^{5} & %
\mbox{\qquad\qquad\rule[-0.4cm]{0cm}{1cm}} & \nabla \times \mathbf{%
\varepsilon }+\partial _{\tau }\mathbf{b}=0%
\end{array}
\\ 
\rule[-0.4cm]{0cm}{1cm}\nabla \varepsilon ^{0}-\partial _{\tau }\mathbf{e}%
+\partial _{0}\mathbf{\varepsilon }=0%
\end{array}
\label{3v}
\end{equation}%
and the Poynting vectors are%
\begin{eqnarray}
T^{00} &=&\frac{\lambda }{2}\left[ \mathbf{e}\cdot \mathbf{e}^{\Phi }+%
\mathbf{b}\cdot \mathbf{b}^{\Phi }-\mathbf{\varepsilon }\cdot \mathbf{%
\varepsilon ^{\Phi }-}\varepsilon ^{0}\varepsilon _{\Phi }^{0}\right] %
\mbox{\qquad}T^{0i}=\lambda \left[ \mathbf{e}\times \mathbf{b}^{\Phi
}-\varepsilon ^{0}\mathbf{\varepsilon ^{\Phi }}\right] ^{i} \\
T^{55} &=&\frac{\lambda }{2}\left[ -\mathbf{e}\cdot \mathbf{e}^{\Phi }+%
\mathbf{b}\cdot \mathbf{b}^{\Phi }+\mathbf{\varepsilon }\cdot \mathbf{%
\varepsilon ^{\Phi }-}\varepsilon ^{0}\varepsilon _{\Phi }^{0}\right] %
\mbox{\qquad}T^{5i}=\lambda \left[ \varepsilon ^{0}\mathbf{e}^{\Phi }+%
\mathbf{\varepsilon }\times \mathbf{b}^{\Phi }\right] ^{i}.
\end{eqnarray}%
From the invariants (\ref{inv-2}) and (\ref{inv-3}) it follows that 
\begin{eqnarray}
-\mathbf{e}\cdot \mathbf{e}+\mathbf{b}\cdot \mathbf{b}=\mathbf{\varepsilon }%
\cdot \mathbf{\varepsilon }-\varepsilon ^{0}\varepsilon ^{0}=0 &\Rightarrow
&\left\{ 
\begin{array}{l}
T^{00}=\lambda \left[ \mathbf{e}\cdot \mathbf{e}^{\Phi }-\mathbf{\varepsilon 
}\cdot \mathbf{\varepsilon ^{\Phi }}\right] \\ 
T^{55}=0\mbox{\qquad\rule{0cm}{0.5cm}}%
\end{array}%
\right. \\
\;\mathbf{e}\times \mathbf{b}=\left( \mathbf{e}\cdot \mathbf{e}\right) 
\mathbf{\hat{\varepsilon}}\mbox{\qquad}\varepsilon ^{0}\mathbf{e}=-\mathbf{%
\varepsilon }\times \mathbf{b} &\Rightarrow &\left\{ 
\begin{array}{l}
T^{0i}=\lambda \left[ \mathbf{e}\cdot \mathbf{e}^{\Phi }-\mathbf{\varepsilon 
}\cdot \mathbf{\varepsilon ^{\Phi }}\right] \mathbf{\hat{\varepsilon}}^{i}
\\ 
T^{5i}=0\mbox{\qquad\rule{0cm}{0.5cm}}%
\end{array}%
\right.
\end{eqnarray}%
where from (\ref{rad-2})%
\begin{equation}
\mathbf{\hat{\varepsilon}}=\frac{\mathbf{\varepsilon }}{\left\vert \mathbf{%
\varepsilon }\right\vert }=\mathbf{\hat{z}}
\end{equation}%
again indicating that the Maxwell Poynting vector is oriented along the
observation vector.

A general plane wave expansion is taken by writing the Fourier transform%
\begin{equation}
f^{\alpha \beta }\left( x,\tau \right) =\frac{1}{\left( 2\pi \right) ^{5}}%
\int d^{4}kd\kappa ~e^{i\left( k\cdot x-\kappa \tau \right) }f^{\alpha \beta
}\left( k,\kappa \right)
\end{equation}%
for which the wave equation (\ref{wave}) in vacuum gives%
\begin{equation}
k^{\alpha }k_{\alpha }=k^{\mu }k_{\mu }-\kappa ^{2}=\mathbf{k}%
^{2}-(k^{0})^{2}-\kappa ^{2}=0
\end{equation}%
so that concatenation of $f^{\alpha \beta }\left( x,\tau \right) $ forces
the photon onto the $\kappa \rightarrow 0$ zero-mass shell. The general
plane wave solution to the 3-vector field equations (\ref{3v}) in terms of
the transverse component of $\mathbf{e}$ and the longitudinal component of $%
\mathbf{\varepsilon }$ was given in \cite{emlf} as 
\begin{equation}
\mathbf{e}=\mathbf{e}_{\perp }-{\frac{\kappa }{k^{0}}}~\mathbf{\varepsilon }%
_{\parallel }\mbox{\qquad}\mathbf{h}={\frac{1}{k^{0}}}\mathbf{k}\times 
\mathbf{e}_{\perp }\mbox{\qquad}\mathbf{\varepsilon }=\mathbf{\varepsilon }%
_{\parallel }+{\frac{\kappa }{k^{0}}}~\mathbf{e}_{\perp }\mbox{\qquad}%
\varepsilon ^{0}={\frac{1}{k^{0}}}\mathbf{k}\cdot \mathbf{\varepsilon }%
_{\parallel }.  \label{plnwv}
\end{equation}%
On the zero-mass shell, the $\mathbf{e}$ and $\mathbf{h}$ fields take the
Maxwell form --- mutually orthogonal with equal amplitude, and both
orthogonal to the propagation vector $\mathbf{k}$ --- and the $\mathbf{%
\varepsilon }$ field decouples from $\mathbf{e}$ and $\mathbf{h}$, becoming
purely longitudinal. In terms of solution (\ref{plnwv}) the components of
the energy-momentum tensor are 
\begin{equation}
\begin{array}{lll}
T^{00}=\lambda \left( \mathbf{e}_{\perp }\cdot \mathbf{e}_{\perp }^{\Phi }-%
\mathbf{\varepsilon }_{\parallel }\cdot \mathbf{\varepsilon _{\parallel
}^{\Phi }}\right) & ~ & T^{0i}=\dfrac{k^{i}}{k^{0}}T^{00} \\ 
~ & ~ & ~ \\ 
T^{55}=\left( \dfrac{\kappa }{k^{0}}\right) ^{2}T^{00} & ~ & T^{5\mu }=%
\dfrac{\kappa }{k^{0}}\dfrac{k^{\mu }}{k^{0}}T^{00}%
\end{array}
\label{T_pw}
\end{equation}%
and we see that $T^{5\alpha }$ vanishes on the zero-mass shell. Comparison
of (\ref{T-rad}) with (\ref{T_pw}) for the $\kappa =0$ case shows that the
radiation field from the arbitrary event has the character of an equilibrium
plane wave solution.

\section{Abraham-Lorentz-Dirac Equation}

In the previous section we examined the field produced by an arbitrarily
evolving event $X^{\mu }\left( \tau \right) $, and in particular studied the
characteristics of its radiation field, identified by the far-field
behavior. In this section, we sharpen the characterization of the radiation
field, following Dirac's analysis \cite{dirac} of the Greens functions, and
obtain the ALD equation by specifically accounting for the field produced by
the event's own evolution.

\subsection{Radiation reaction}

The Lorentz force (\ref{Lor}) describes the motion of an event under the
influence of a general electromagnetic field $f^{\alpha \beta }\left( x,\tau
\right) $. Depending on the known boundary conditions, $f^{\alpha \beta
}\left( x,\tau \right) $ may be described as an initial incoming field and
the retarded reaction, or as a final outgoing field and the advanced
reaction, according to 
\begin{equation}
f^{\alpha \beta }\left( x,\tau \right) =f_{in}^{\alpha \beta }\left( x,\tau
\right) +f_{ret}^{\alpha \beta }\left( x,\tau \right) =f_{out}^{\alpha \beta
}\left( x,\tau \right) +f_{adv}^{\alpha \beta }\left( x,\tau \right)
\label{bc}
\end{equation}%
where%
\begin{equation}
f_{ret}^{\alpha \beta }\left( x,\tau \right) \underset{\tau \rightarrow
-\infty }{\xrightarrow{\hspace*{1cm}}}0\mbox{\qquad}\mbox{\qquad}%
f_{adv}^{\alpha \beta }\left( x,\tau \right) \underset{\tau \rightarrow
\infty }{\xrightarrow{\hspace*{1cm}}}0.
\end{equation}%
Thus, when an incoming field $f_{in}^{\alpha \beta }\left( x,\tau \right) $
impinges on the event, the Lorentz force can be expressed as%
\begin{equation}
m\ddot{X}^{\mu }\left( \tau \right) =\lambda e\left[ f_{in}^{\mu \beta
}\left( X,\tau \right) +f_{ret}^{\mu \beta }\left( X,\tau \right) \right] 
\dot{X}_{\beta }\left( \tau \right)  \label{Lor-2}
\end{equation}%
where $f_{ret}^{\alpha \beta }\left( x,\tau \right) $ is produced by the
event $X^{\mu }\left( \tau \right) $ itself through the field equation (\ref%
{pMm}). However, the interaction of the event with its own field includes an
infinite part, and to identify the finite part, Dirac argued that the
outgoing field can be expressed as the sum of the incoming field and a
radiation reaction produced when the incoming field accelerates the event.
Thus,%
\begin{equation}
f_{out}^{\alpha \beta }=f_{in}^{\alpha \beta }+f_{rad}^{\alpha \beta }
\label{out-def}
\end{equation}%
which combines with (\ref{bc}) to provide%
\begin{equation}
f_{rad}^{\alpha \beta }=f_{out}^{\alpha \beta }-f_{in}^{\alpha \beta
}=\left( f^{\alpha \beta }-f_{adv}^{\alpha \beta }\right) -\left( f^{\alpha
\beta }-f_{ret}^{\alpha \beta }\right) =f_{ret}^{\alpha \beta
}-f_{adv}^{\alpha \beta }~.  \label{r-a}
\end{equation}%
Considering the combinations%
\begin{equation}
f_{ret}^{\alpha \beta }=\frac{1}{2}\left( f_{ret}^{\alpha \beta
}-f_{adv}^{\alpha \beta }\right) +\frac{1}{2}\left( f_{ret}^{\alpha \beta
}+f_{adv}^{\alpha \beta }\right) =\frac{1}{2}f_{rad}^{\alpha \beta }+\frac{1%
}{2}\left( f_{ret}^{\alpha \beta }+f_{adv}^{\alpha \beta }\right)
\end{equation}%
the radiation reaction $f_{rad}^{\alpha \beta }$ is found to be finite,
while the the infinite part of the self-interaction $f_{ret}^{\alpha \beta
}+f_{adv}^{\alpha \beta }$ is treated by mass renormalization. The finite
part of the Lorentz force is then%
\begin{equation}
m\ddot{X}^{\mu }=\lambda e\left[ \frac{1}{2}f_{rad}^{\mu \beta }+\
f_{in}^{\mu \beta }\right] \dot{X}_{\beta }~.  \label{Lor-finite}
\end{equation}%
From the prescription (\ref{r-a}) the potential $a_{rad}^{\beta }\left(
x,\tau \right) $ is found by making the substitution%
\begin{equation}
\theta ^{ret}\longrightarrow \theta ^{rad}=\theta ^{ret}-\theta
^{adv}=\epsilon \left( x^{0}-X^{0}\right)
\end{equation}%
in (\ref{a_usefl}), so that the self-interaction term in (\ref{Lor-finite})
is the force%
\begin{equation}
F_{\emph{self}}^{\mu }\left( \tau \right) =\frac{1}{2}\lambda ef_{rad}^{\mu
\beta }\left( X\left( \tau \right) ,\tau \right) \dot{X}_{\beta }\left( \tau
\right)  \label{Fsf}
\end{equation}%
evaluated at the event $X^{\mu }\left( \tau \right) $. The field must
therefore be calculated from%
\begin{equation}
\partial ^{\mu }a_{rad}^{\beta }\left( x,\tau \right) =-{\frac{e}{{2\pi }}}%
\int ds\frac{d}{ds}\left[ \varphi (\tau -s)\frac{\dot{X}^{\beta }\left(
s\right) \left[ x^{\mu }-X^{\mu }\left( s\right) \right] }{\dot{X}\left(
s\right) \cdot \left[ x-X\left( s\right) \right] }\right] \theta
^{rad}\delta \left[ \left( x-X\left( s\right) \right) ^{2}\right]
\label{a-rad}
\end{equation}%
in the limit%
\begin{equation}
x^{\mu }-X^{\mu }\left( s\right) \rightarrow X^{\mu }\left( \tau \right)
-X^{\mu }\left( s\right)
\end{equation}%
which by conditions (\ref{z-dot}) and (\ref{z-sq}) require%
\begin{equation}
\left[ X\left( \tau \right) -X\left( s\right) \right] ^{2}=0\mbox{\qquad}%
\Rightarrow \mbox{\qquad}s\rightarrow \tau .
\end{equation}%
Dirac showed \cite{dirac} that expressions of the type (\ref{a-rad}) have a
finite limit as $X^{\mu }\left( s\right) \rightarrow X^{\mu }\left( \tau
\right) $, found by performing a Taylor expansion along worldline $X\left(
\tau \right) $, that is, evaluating%
\begin{equation}
\partial ^{\mu }a_{rad}^{\beta }\left( X\left( \tau ^{\prime }\right) ,\tau
\right)
\end{equation}%
where 
\begin{eqnarray}
\tau ^{\prime } &=&\tau +h\mbox{\rule[-0.5cm]{0cm}{1.0cm}} \\
X\left( \tau ^{\prime }\right) -X\left( \tau \right) &=&h\dot{X}\left( \tau
\right) +\frac{h^{2}}{2}\ddot{X}\left( \tau \right) +\frac{h^{3}}{3!}\dddot{X%
}\left( \tau \right) +\cdots \mbox{\rule[-0.5cm]{0cm}{1.0cm}} \\
\dot{X}\left( \tau ^{\prime }\right) &=&\dot{X}\left( \tau \right) +h\ddot{X}%
\left( \tau \right) +\frac{h^{2}}{2}\dddot{X}\left( \tau \right) +\cdots
\end{eqnarray}%
By comparison of (\ref{a-rad}) with the expression for the field in standard
Maxwell theory \cite{dirac,barut}%
\begin{equation}
\partial ^{\mu }A_{rad}^{\nu }\left( x\right) =-{\frac{e}{{2\pi }}}\int ds%
\frac{d}{ds}\left[ \frac{\dot{X}^{\nu }\left( s\right) \left[ x^{\mu
}-X^{\mu }\left( s\right) \right] }{\dot{X}\left( s\right) \cdot \left[
x-X\left( s\right) \right] }\right] \theta ^{rad}\delta \left[ \left(
x-X\left( s\right) \right) ^{2}\right] ~,
\end{equation}%
the field expansion in off-shell electrodynamics is seen to take the
standard form%
\begin{equation}
f_{rad}^{\mu \nu }\left( X\left( \tau \right) ,\tau \right) =\frac{2}{3}%
\frac{e}{2\pi }\left[ \dot{X}^{\mu }\left( \tau \right) \dddot{X}^{\nu
}\left( \tau \right) -\dot{X}^{\nu }\left( \tau \right) \dddot{X}^{\mu
}\left( \tau \right) \right]  \label{self-force}
\end{equation}%
where we use $\varphi \left( 0\right) =\frac{1}{2\lambda }$.%
\QTR{framesubtitle}{\ Now, the self-interaction term in (\ref{Fsf}) }is%
\begin{equation}
F_{\emph{self}}^{\mu }\left( \tau \right) =\lambda \frac{2}{3}\frac{e^{2}}{%
2\pi }\left[ \dot{X}^{\mu }\dddot{X}^{\nu }-\dot{X}^{\nu }\dddot{X}^{\mu }%
\right] \dot{X}_{\nu }=\frac{2}{3}\frac{e^{2}}{4\pi }\left[ \dot{X}^{\mu }%
\dddot{X}^{\nu }\dot{X}_{\nu }-\dot{X}^{2}\dddot{X}^{\mu }\right] .
\end{equation}%
Since the radiation field was seen to carry no mass, the event may be
considered on-shell, so that%
\begin{equation}
\dot{X}^{2}=-1\mbox{\qquad}\dot{X}\cdot \ddot{X}=0\mbox{\qquad}\dot{X}\cdot 
\dddot{X}+\ddot{X}^{2}=0,
\end{equation}%
and the \QTR{framesubtitle}{Lorentz force} (\ref{Lor-finite}) takes the form
of the usual ALD equation%
\begin{equation}
m\ddot{x}^{\mu }=e_{0}f_{in}^{\mu \beta }\left( x,\tau \right) \dot{x}%
_{\beta }+m\tau _{0}\left[ \dddot{x}^{\mu }-\ddot{x}^{2}~\dot{x}^{\mu }%
\right]  \label{ALD-2}
\end{equation}%
where we have written $\tau _{0}$ as in (\ref{t0}), and reverted to lower
case for the event trajectory 
\begin{equation}
X^{\mu }\left( \tau \right) \rightarrow x^{\mu }\left( \tau \right)
\end{equation}
since the field point is understood to be the event location%
\begin{equation}
f_{in}^{\mu \beta }\left( x,\tau \right) =f_{in}^{\mu \beta }\left( x\left(
\tau \right) ,\tau \mbox{\rule{0cm}{10.5pt}}\right) .
\end{equation}

\subsection{Integro-differential equation}

In order to eliminate the run-away solutions admitted by the ALD equation in
the form (\ref{ALD-2}), we convert it to an integro-differential equation 
\begin{eqnarray}
\left( \ddot{x}^{\mu }-\tau _{0}\dddot{x}^{\mu }\right) e^{-\tau /\tau _{0}}
&&\mbox{\hspace{-20pt}}=e^{-\tau /\tau _{0}}\left[ \frac{e_{0}}{m}%
f_{in}^{\mu \beta }\left( x,\tau \right) \dot{x}_{\beta }-\tau _{0}\ddot{x}%
^{2}~\dot{x}^{\mu }\right] \mbox{\rule[-0.5cm]{0cm}{1.0cm}} \\
\ddot{x}^{\mu }e^{-\tau /\tau _{0}} &&\mbox{\hspace{-20pt}}=-\int_{0}^{\tau
}ds~e^{-s/\tau _{0}}\left[ \frac{e_{0}}{\tau _{0}m}f_{in}^{\mu \beta }\left(
x,s\right) \dot{x}_{\beta }-\ddot{x}^{2}~\dot{x}^{\mu }\right] +\ddot{x}%
^{\mu }\left( 0\right) .  \label{i-1}
\end{eqnarray}%
The run-away solutions are suppressed by imposing the requirement that
velocity grow less than exponentially for long times%
\begin{equation}
\ddot{x}^{\mu }\left( \tau \right) e^{-\tau /\tau _{0}}\underset{\tau
\rightarrow \infty }{\xrightarrow{\hspace*{1cm}}}0  \label{BC}
\end{equation}%
so that as $\tau \rightarrow \infty $ (\ref{i-1}) becomes%
\begin{equation}
0=-\int_{0}^{\infty }ds~e^{-s/\tau _{0}}\left[ \frac{e_{0}}{\tau _{0}m}%
f_{in}^{\mu \beta }\left( x,s\right) \dot{x}_{\beta }\left( s\right) -\ddot{x%
}^{2}\left( s\right) \dot{x}^{\mu }\left( s\right) \right] +\ddot{x}^{\mu
}\left( 0\right)   \label{BC-1}
\end{equation}%
which is incorporated back into (\ref{i-1}) to provide the ALD equation as%
\begin{equation}
m\ddot{x}^{\mu }\left( \tau \right) =\int_{\tau }^{\infty }ds~\frac{1}{\tau
_{0}}e^{-\left( s-\tau \right) /\tau _{0}}\left[ e_{0}f_{in}^{\mu \beta
}\left( x\left( s\right) \mbox{\rule{0cm}{10.5pt}},s\right) \dot{x}_{\beta
}\left( s\right) -m\tau _{0}\ddot{x}^{2}\left( s\right) \dot{x}^{\mu }\left(
s\right) \right] .  \label{ALD-3}
\end{equation}%
This expression describes an acceleration at $\tau $ that depends on values
of the interaction at later times, as was seen for the standard ALD equation
(\ref{ALDI}), which is recovered from (\ref{ALD-3}) in the Maxwell limit%
\begin{equation}
e_{0}f_{in}^{\mu \beta }\left( x\left( \tau \right) ,\tau %
\mbox{\rule{0cm}{10.5pt}}\right) =e\lambda f_{in}^{\mu \beta }\left( x\left(
\tau \right) ,\tau \mbox{\rule{0cm}{10.5pt}}\right) \rightarrow eF_{in}^{\mu
\beta }\left( x\left( \tau \right) \mbox{\rule{0cm}{10.5pt}}\right) .
\end{equation}%
In the context of Maxwell theory, the field depends only on the spacetime
location of the particle on which it acts, so that the synchronization of
the particle-field interaction is sharply determined by the proper time $%
\tau $ of the particle motion. %
%
In the off-shell formalism, however, the $\tau $-dependence of the
electromagnetic field typically introduces a statistical synchronization
between interacting events, on a time scale $\lambda $ much larger than $%
\tau _{0}$. The integration in (\ref{ALD-3}) may thus be understood as
analogous to this smoothed synchronization, describing a modified
self-interaction between the event and a radiation field induced by an
ensemble associated with the event evolution.

To clarify this interpretation, we use (\ref{EuM}) to express $f_{in}^{\mu
\beta }\left( x,\tau \right) $ in terms of $f_{\Phi }^{\mu \beta }\left(
x,\tau \right) $, a sharp field induced by a sharp event density $j\left(
x,\tau \right) $ of the type (\ref{pM}). From (\ref{inverse}) and (\ref{pMm}%
) the external field can be written as an ensemble average of such sharp
fields%
\begin{equation}
f_{in}^{\mu \beta }\left( x,\tau \right) =\int_{-\infty }^{\infty
}ds~f_{\Phi }^{\mu \beta }\left( x\left( \tau \right) ,s\mbox{%
\rule{0cm}{10.5pt}}\right) \varphi \left( s-\tau \right)   \label{F_shrp}
\end{equation}%
leading to an integral expression for the Lorentz force (\ref{Lor})%
\begin{equation}
m\ddot{x}^{\mu }\left( \tau \right) =e_{0}f^{\mu \beta }\left( x\left( \tau
\right) ,\tau \mbox{\rule{0cm}{10.5pt}}\right) \dot{x}_{\beta }\left( \tau
\right) =\int_{-\infty }^{\infty }ds~e_{0}f_{\Phi }^{\mu \beta }\left(
x\left( \tau \right) ,s\mbox{\rule{0cm}{10.5pt}}\right) \varphi \left(
s-\tau \right) \dot{x}_{\beta }\left( \tau \right) .  \label{Lor-shrp}
\end{equation}%
By comparison, the ALD equation (\ref{ALD-3}) can be written as%
\begin{equation}
m\ddot{x}^{\mu }\left( \tau \right) =\int_{-\infty }^{\infty }ds~\left[
e_{0}f_{in}^{\mu \beta }\left( x\left( s\right) ,s\mbox{\rule{0cm}{10.5pt}}%
\right) \dot{x}_{\beta }\left( s\right) -m\tau _{0}\ddot{x}^{2}\left(
s\right) \dot{x}^{\mu }\left( s\right) \right] \phi \left( s-\tau \right) 
\label{ALD-3a}
\end{equation}%
where we interpret the integrating factor as a normalized distribution%
\begin{equation}
\phi \left( \tau \right) =\frac{1}{\tau _{0}}e^{-\tau /\tau _{0}}\theta
\left( \tau \right) \mbox{\qquad}\int_{-\infty }^{\infty }d\tau ~\phi \left(
\tau \right) =1.  \label{d-2}
\end{equation}%
Aside from the additional interaction term $\ddot{x}^{2}\dot{x}^{\mu }$,
equation (\ref{ALD-3a}) differs from (\ref{Lor-shrp}) in that the
statistical synchronization affected by the distribution $\phi \left( \tau
\right) $ extends to both the field $f_{in}^{\mu \beta }\left( x,s\right) $
and the velocity $\dot{x}_{\beta }\left( s\right) $.

In order to examine the causal structure of (\ref{ALD-3}) more closely, we
shift the integration variable as $s\rightarrow s-\tau $%
\begin{eqnarray}
m\ddot{x}^{\mu }\left( \tau \right)  &&\mbox{\hspace{-20pt}}%
=e_{0}\int_{0}^{\infty }ds~\frac{1}{\tau _{0}}e^{-s/\tau _{0}}~f_{in}^{\mu
\beta }\left( x,s+\tau \right) \dot{x}_{\beta }\left( s+\tau \right)   \notag
\\
&&-m\tau _{0}\int_{0}^{\infty }ds~\frac{1}{\tau _{0}}e^{-s/\tau _{0}}~\ddot{x%
}^{2}\left( s+\tau \right) \dot{x}^{\mu }\left( s+\tau \right) .
\end{eqnarray}%
and consider the genealogy of the field. Although the incoming field $%
f_{in}^{\mu \beta }\left( x,\tau \right) $ is taken to be external, it must
have been induced by the smoothed current $j_{\varphi }\left( x,\tau \right) 
$ associated with the sharp event densities $j\left( x,\tau \right) $
distributed along the worldlines of some configuration of evolving events.
Using (\ref{F_shrp}) for the field induced directly by the sharp event
density $j\left( x,\tau \right) $, we rewrite (\ref{ALD-3}) as%
\begin{eqnarray}
m\ddot{x}^{\mu }\left( \tau \right)  &&\mbox{\hspace{-20pt}}%
=e_{0}\int_{0}^{\infty }ds~\frac{1}{\tau _{0}}e^{-s/\tau _{0}}\int_{-\infty
}^{\infty }ds^{\prime }~\frac{1}{2\lambda }e^{-\left\vert s+\tau -s^{\prime
}\right\vert /\lambda }~f_{\Phi }^{\mu \beta }\left( x,s^{\prime }\right) 
\dot{x}_{\beta }\left( s+\tau \right)   \notag \\
&&-m\tau _{0}\int_{0}^{\infty }ds~\frac{1}{\tau _{0}}e^{-s/\tau _{0}}~\ddot{x%
}^{2}\left( s+\tau \right) \dot{x}^{\mu }\left( s+\tau \right) 
\label{ALD-4}
\end{eqnarray}%
and rearrange the first line by combining the exponential functions%
\begin{equation}
e^{-s/\tau _{0}}e^{-\left\vert s+\tau -s^{\prime }\right\vert /\lambda
}=\left\{ 
\begin{array}{lll}
\exp \left[ -s\left( \frac{1}{\tau _{0}}+\frac{1}{\lambda }\right) -\left(
\tau -s^{\prime }\right) \frac{1}{\lambda }\right]  & ,\mbox{%
\rule[-0.5cm]{0cm}{1.0cm}} & -\infty <s^{\prime }\leq s+\tau  \\ 
\exp \left[ -s\left( \frac{1}{\tau _{0}}-\frac{1}{\lambda }\right) +\left(
\tau -s^{\prime }\right) \frac{1}{\lambda }\right]  & , & s+\tau \leq
s^{\prime }<\infty 
\end{array}%
\right. 
\end{equation}%
and partitioning the $s^{\prime }$-integration according to%
\begin{equation}
\int_{0}^{\infty }ds\int_{-\infty }^{\infty }ds^{\prime }~\left\{ \cdots
\right\} =\int_{0}^{\infty }ds\left[ \int_{-\infty }^{s+\tau }ds^{\prime
}\left\{ \cdots \right\} +\int_{s+\tau }^{\infty }ds^{\prime }\left\{ \cdots
\right\} \right] .
\end{equation}%
The relationship of time scales $\lambda \gg \tau _{0}$ permits the
approximation%
\begin{equation}
\exp \left[ -s\left( \frac{1}{\tau _{0}}\pm \frac{1}{\lambda }\right) \right]
=e^{-s/\tau _{0}}+o\left( \frac{\tau _{0}}{\lambda }\right) 
\end{equation}%
making no assumptions about the functional forms of events or fields, and so
the integrals in the first line of (\ref{ALD-3}) becomes%
\begin{eqnarray}
\int_{0}^{\infty } &&\mbox{\hspace{-20pt}}ds\frac{1}{\tau _{0}}e^{-s/\tau
_{0}}~\dot{x}_{\beta }\left( s+\tau \right) \times \mbox{%
\rule[-0.5cm]{0cm}{1.0cm}}  \notag \\
&&\left\{ \int_{-\infty }^{s+\tau }ds^{\prime }~\frac{1}{2\lambda }%
e^{-\left( \tau -s^{\prime }\right) /\lambda }~f_{\Phi }^{\mu \beta }\left(
x,s^{\prime }\right) +\int_{s+\tau }^{\infty }ds^{\prime }~\frac{1}{2\lambda 
}e^{\left( \tau -s^{\prime }\right) /\lambda }~f_{\Phi }^{\mu \beta }\left(
x,s^{\prime }\right) \right\} .
\end{eqnarray}%
Further partitioning the $s^{\prime }$-integrations as%
\begin{eqnarray}
\int_{-\infty }^{s+\tau }ds^{\prime } &&~\mbox{\hspace{-20pt}}\frac{1}{%
2\lambda }e^{-\left( \tau -s^{\prime }\right) /\lambda }~f_{\Phi }^{\mu
\beta }\left( x,s^{\prime }\right) =\int_{-\infty }^{\tau }ds^{\prime }~%
\frac{1}{2\lambda }e^{-\left\vert \tau -s^{\prime }\right\vert /\lambda
}~f_{\Phi }^{\mu \beta }\left( x,s^{\prime }\right) \mbox{%
\rule[-0.5cm]{0cm}{1.0cm}}  \notag \\
&&\mbox{\hspace{3.5cm}}+\int_{\tau }^{s+\tau }ds^{\prime }~\frac{1}{2\lambda 
}e^{-\left( \tau -s^{\prime }\right) /\lambda }~f_{\Phi }^{\mu \beta }\left(
x,s^{\prime }\right) 
\end{eqnarray}%
and%
\begin{eqnarray}
\int_{s+\tau }^{\infty }ds^{\prime } &&~\mbox{\hspace{-20pt}}\frac{1}{%
2\lambda }e^{\left( \tau -s^{\prime }\right) /\lambda }~f_{\Phi }^{\mu \beta
}\left( x,s^{\prime }\right) =\int_{\tau }^{\infty }ds^{\prime }~\frac{1}{%
2\lambda }e^{-\left\vert \tau -s^{\prime }\right\vert /\lambda }~f_{\Phi
}^{\mu \beta }\left( x,s^{\prime }\right) \mbox{\rule[-0.5cm]{0cm}{1.0cm}} 
\notag \\
&&\mbox{\hspace{3.5cm}}-\int_{\tau }^{s+\tau }ds^{\prime }~\frac{1}{2\lambda 
}e^{\left( \tau -s^{\prime }\right) /\lambda }~f_{\Phi }^{\mu \beta }\left(
x,s^{\prime }\right) 
\end{eqnarray}%
the integrals in the first line of (\ref{ALD-3}) are now%
\begin{eqnarray}
&&\mbox{\hspace{-20pt}}\int_{0}^{\infty }ds\frac{1}{\tau _{0}}e^{-s/\tau
_{0}}~\dot{x}_{\beta }\left( s+\tau \right) \times \mbox{%
\rule[-0.5cm]{0cm}{1.0cm}}  \notag \\
&&\mbox{\qquad}\left\{ \int_{-\infty }^{\infty }ds^{\prime }~\frac{1}{%
2\lambda }e^{-\left\vert \tau -s^{\prime }\right\vert /\lambda }~f_{\Phi
}^{\mu \beta }\left( x,s^{\prime }\right) \right.   \notag \\
&&\mbox{\qquad}\mbox{\qquad}+\left. \int_{\tau }^{s+\tau }ds^{\prime }~\frac{%
1}{2\lambda }\left[ e^{-\left( \tau -s^{\prime }\right) /\lambda }-e^{\left(
\tau -s^{\prime }\right) /\lambda }\right] ~f_{\Phi }^{\mu \beta }\left(
x,s^{\prime }\right) \right\} +o\left( \frac{\tau _{0}}{\lambda }\right) 
\label{line-2}
\end{eqnarray}%
in which, from (\ref{F_shrp}), we recognize the second line as $f_{in}^{\mu
\beta }\left( x,\tau \right) $. Shifting $s^{\prime }-\tau \rightarrow
s^{\prime }/\lambda $ in the third line of (\ref{line-2}) leads to%
\begin{equation}
\int_{\tau }^{s+\tau }ds^{\prime }~\frac{1}{\lambda }\sinh \left( \frac{%
s^{\prime }-\tau }{\lambda }\right) f_{\Phi }^{\mu \beta }\left( x,s^{\prime
}\right) =\int_{0}^{s/\lambda }ds^{\prime }~\sinh \left( s^{\prime }\right)
~f_{\Phi }^{\mu \beta }\left( x,\lambda s^{\prime }+\tau \right) 
\end{equation}%
which for smooth $f_{\Phi }$ may be estimated as being of order $o\left(
\tau _{0}^{2}/\lambda ^{2}\right) $:%
\begin{eqnarray}
\int_{0}^{s/\lambda }ds^{\prime }~\sinh \left( s^{\prime }\right) f_{\Phi
}^{\mu \beta }\left( x,\lambda s^{\prime }+\tau \right)  &&%
\mbox{\hspace{-20pt}}\simeq \int_{0}^{\tau _{0}/\lambda }ds^{\prime }~\sinh
\left( s^{\prime }\right) ~f_{\Phi }^{\mu \beta }\left( x,\lambda s^{\prime
}+\tau \right) \mbox{\rule[-0.5cm]{0cm}{1.0cm}} \\
&&\mbox{\hspace{-20pt}}\simeq \frac{\tau _{0}^{2}}{\lambda ^{2}}f_{\Phi
}^{\mu \beta }\left( x,\tau \right) -\frac{\tau _{0}^{3}}{3\lambda ^{2}}%
\frac{d}{d\tau }f_{\Phi }^{\mu \beta }\left( x,\tau \right) .  \label{est}
\end{eqnarray}%
Shifting the $s$-integral back to $s\rightarrow s-\tau $, and using the
distribution $\phi \left( \tau \right) $ defined in (\ref{d-2}) to extend
the limits of integration, the ALD equation (\ref{ALD-3}) now takes the form%
\begin{eqnarray}
m\ddot{x}^{\mu }\left( \tau \right)  &&\mbox{\hspace{-20pt}}%
=e_{0}\int_{-\infty }^{\infty }ds~f_{in}^{\mu \beta }\left( x\left( s\right)
,\tau \mbox{\rule{0cm}{10.5pt}}\right) \phi \left( s-\tau \right) ~\dot{x}%
_{\beta }\left( s\right) +o\left( \frac{\tau _{0}}{\lambda }\right) %
\mbox{\rule[-0.5cm]{0cm}{1.0cm}}  \notag \\
&&\mbox{\hspace{-20pt}}+e_{0}\int_{-\infty }^{\infty }ds~\phi \left( s-\tau
\right) ~\dot{x}_{\beta }\left( s\right) \int_{0}^{\left( s-\tau \right)
/\lambda }ds^{\prime }~\sinh \left( s^{\prime }\right) f_{\Phi }^{\mu \beta
}\left( x\left( s\right) ,\lambda s^{\prime }+\tau \mbox{\rule{0cm}{10.5pt}}%
\right) \mbox{\rule[-0.5cm]{0cm}{1.0cm}}  \notag \\
&&\mbox{\hspace{-20pt}}-m\tau _{0}\int_{-\infty }^{\infty }ds~\phi \left(
s-\tau \right) ~\ddot{x}^{2}\left( s\right) \dot{x}^{\mu }\left( s\right) 
\label{ALD-6}
\end{eqnarray}%
where we have made no assumptions about the $\tau $-dependence of events or
fields. In this expression, the acceleration at time $\tau $ is given by the
interaction of the field at time $\tau $ with a narrow ensemble of events
described by the distribution $\phi \left( \tau \right) $. 

If the field $f_{\Phi }^{\mu \beta }\left( x,\tau \right) $ varies smoothly,
we see from (\ref{est}) that the second line in (\ref{ALD-6}) can be
neglected. However, this term may be significant for external fields of the
type (\ref{f-1}), which are of the form%
\begin{equation}
f_{in}^{\mu \beta }\left( x,\tau \right) =\varphi \left( \tau -\tau
_{ret}\right) \left. F_{Maxwell}^{\mu \beta }\left( x\right) \right\vert
_{\tau _{ret}}
\end{equation}%
and associated with the sharp form%
\begin{equation}
f_{\Phi }^{\mu \beta }\left( x,\tau \right) =\delta \left( \tau -\tau
_{ret}\right) \left. F_{Maxwell}^{\mu \beta }\left( x\right) \right\vert
_{\tau _{ret}}~.
\end{equation}%
Then, the integral $s^{\prime }$ integration is%
\begin{eqnarray}
\int_{0}^{\left( s-\tau \right) /\lambda }ds^{\prime }\sinh \left( s^{\prime
}\right) f_{\Phi }^{\mu \beta }\left( x,\lambda s^{\prime }+\tau \right)  &&%
\mbox{\hspace{-20pt}}=F_{Maxwell}^{\mu \beta }\left( x\right) \times   \notag
\\
&&\mbox{\hspace{-20pt}}\int_{0}^{\left( s-\tau \right) /\lambda }ds^{\prime
}~\sinh \left( s^{\prime }\right) \delta \left( \lambda s^{\prime }+\tau
-\tau _{ret}\right)  \\
=F_{Maxwell}^{\mu \beta }\left( x\right)  &&\mbox{\hspace{-20pt}}\sinh
\left( \frac{\tau -\tau _{ret}}{\lambda }\right) \theta \left( s-\tau
_{ret}\right) 
\end{eqnarray}%
and the $s$ integration becomes%
\begin{equation}
e_{0}\int_{\tau _{ret}}^{\infty }ds~\frac{1}{\tau _{0}}e^{-\left( s-\tau
\right) /\tau _{0}}~F_{Maxwell}^{\mu \beta }\left( x\left( s\right) %
\mbox{\rule{0cm}{10.5pt}}\right) \dot{x}_{\beta }\left( s\right) \sinh
\left( \frac{\tau -\tau _{ret}}{\lambda }\right) 
\end{equation}%
which can be put into the form%
\begin{eqnarray}
e_{0} &&\mbox{\hspace{-20pt}}e^{-\left( \tau -\tau _{ret}\right) /\tau
_{0}}\sinh \left( \frac{\tau -\tau _{ret}}{\lambda }\right) \theta \left(
\tau -\tau _{ret}\right) \times  \\
&&\int_{\tau _{ret}}^{\infty }ds~\frac{1}{\tau _{0}}~F_{Maxwell}^{\mu \beta
}\left( x\left( s\right) \mbox{\rule{0cm}{10.5pt}}\right) \dot{x}_{\beta
}\left( s\right) e^{-\left( s-\tau _{ret}\right) /\tau _{0}}
\end{eqnarray}%
so that the function 
\begin{equation}
\psi \left( \tau -\tau _{ret}\right) =e^{-\left( \tau -\tau _{ret}\right)
/\tau _{0}}\sinh \left( \frac{\tau -\tau _{ret}}{\lambda }\right) \theta
\left( \tau -\tau _{ret}\right) \sim o\left( \frac{\tau _{0}}{\lambda }%
\right) 
\end{equation}%
limits the significance of this term. For fields of this type, ALD takes the
form%
\begin{eqnarray}
m\ddot{x}^{\mu }\left( \tau \right)  &&\mbox{\hspace{-20pt}}%
=e_{0}\int_{-\infty }^{\infty }ds~f_{in}^{\mu \beta }\left( x\left( s\right)
,\tau \mbox{\rule{0cm}{10.5pt}}\right) \phi \left( s-\tau \right) ~\dot{x}%
_{\beta }\left( s\right) \mbox{\rule[-0.5cm]{0cm}{1.0cm}}  \notag \\
&&\mbox{\hspace{-20pt}}+e_{0}\psi \left( \tau -\tau _{ret}\right)
\int_{-\infty }^{\infty }ds~\phi \left( s-\tau _{ret}\right)
~F_{Maxwell}^{\mu \beta }\left( x\left( s\right) \mbox{\rule{0cm}{10.5pt}}%
\right) \dot{x}_{\beta }\left( s\right) \mbox{\rule[-0.5cm]{0cm}{1.0cm}} 
\notag \\
&&\mbox{\hspace{-20pt}}-m\tau _{0}\int_{-\infty }^{\infty }ds~\phi \left(
s-\tau \right) ~\ddot{x}^{2}\left( s\right) \dot{x}^{\mu }\left( s\right)
+o\left( \frac{\tau _{0}}{\lambda }\right) .  \label{ALD-7}
\end{eqnarray}%
The ALD equations, in the form (\ref{ALD-6}) or (\ref{ALD-7}), now appear as
an interaction between the instantaneous external field and short-range
ensemble averages over the specific combinations of event velocity $\dot{x}%
^{\mu }\left( \tau \right) $ and acceleration $\ddot{x}^{\mu }\left( \tau
\right) $ that produce the radiation field (\ref{self-force}). These
averages smooth the $\tau $-synchronization between the event and its
radiation reaction, an effect that is qualitatively analogous to the
statistical synchronization expressed in (\ref{pMm}) between the event and
its induced current.

\section{Conclusion}

Historically, the Abraham-Lorentz equation, a nonrelativistic approximation
to (\ref{ALD}), was first obtained \cite{abraham, lorentz} by adding an
effective term to the Lorentz force to account for the energy lost by an
accelerating particle to Larmor radiation. In the $\left\vert \mathbf{\dot{x}%
}\right\vert \ll 1$ approximation (in which case $\tau \rightarrow t$), ALD
reduces to%
\begin{equation}
\mathbf{\ddot{x}}\left( t\right) -\tau _{0}\mathbf{\dddot{x}}\left( t\right)
=\frac{e}{m}\mathbf{E}_{ext}\left( t,\mathbf{x}\right)  \label{AL}
\end{equation}%
which admits a runaway solution for $\mathbf{E}_{ext}=0$ given by%
\begin{equation}
\mathbf{\dot{x}}=\mathbf{\dot{x}}\left( 0\right) e^{t/\tau _{0}}.
\label{runaway}
\end{equation}%
Imposing once again the boundary condition (\ref{BC}), which merely requires
that velocity grow less than exponentially over long times, equation (\ref%
{AL}) may be converted to the integro-differential equation 
\begin{equation}
\mathbf{\ddot{x}}\left( t\right) =\frac{e}{m}\int_{-\infty }^{\infty
}dt^{\prime }~\left[ \frac{1}{\tau _{0}}e^{-\left( t^{\prime }-t\right)
/\tau _{0}}\theta \left( t^{\prime }-t\right) \right] ~\mathbf{E}%
_{ext}\left( t^{\prime },\mathbf{x}\left( t^{\prime }\right) %
\mbox{\rule{0cm}{10.5pt}}\right) ,  \label{AL-1}
\end{equation}%
suppressing the spontaneous acceleration. However, the seemingly innocent
boundary condition, without which the integration over $t^{\prime }$ could
not be made sensible, provides just the pretext under which the
pre-acceleration evident in (\ref{AL-1}) is granted admissibility.
Conversely, to insist that one cannot give meaning to the apparent violation
of classical causality under the integral is equivalent to rejecting the
reasonableness of the boundary condition. Jackson \cite{jackson} summarizes
the conventional interpretation of this situation by accepting a possible
violation of microscopic causality over time scale $\tau _{0}$, because it
cannot disagree with experiment. On this view, since the external field
cannot undergo a macroscopic binary transition in an interval smaller than $%
\tau _{0}$, and since measurements at these time scales will be dominated by
quantum effects, the violation of classical causality in classical mechanics
leads to no experimental contradiction. Although the relativistic ALD
equation (\ref{ALDI}) may be accepted with the same qualification, an
interpretation that preserves some sense of retarded causality in the
classical context would be more satisfying.

As derived in the off-shell electrodynamics associated with Stueckelberg
canonical covariant mechanics, the ALD equation in the form (\ref{ALD-3})
was seen to differ from the standard expression in Maxwell theory (\ref{ALDI}%
) only in the dependence of the external field on the invariant time
parameter $\tau $. It was shown in section 2 that the general $\tau $%
-dependence of off-shell fields can be understood as an expansion of the
U(1) local gauge group to include the invariant evolution parameter, but in
any case, follows directly from the unconstrained commutation relations (\ref%
{com-rel}) among coordinates and velocities of spacetime events. Thus,
off-shell electrodynamics describes a microscopic interaction between
spacetime events $x^{\mu }\left( \tau \right) $ mediated by instantaneous $%
\tau $-dependent fields $f_{\alpha \beta }\left( x,\tau \right) $. However,
it was seen in (\ref{pMm}) that while the off-shell fields are $\tau $%
-dependent, they are induced by an event current associated with an ensemble
of events distributed in $\tau $ along the particle worldline. This
distribution introduces an underlying statistical structure to the classical
event-event interaction, equivalent to relaxing the sharp $\tau $%
-synchronization found in expressions (\ref{ALDI}) and (\ref{AL-1}) between
an event and a field with which it interacts. Detailed consideration of the
resulting $\tau $-dependence of the external field enables the
transformation of equation (\ref{ALD-3}) to the form (\ref{ALD-6}), in which
the Lorentz force depends on the instantaneous value of the external field
and short-range ensemble averages over combinations of event velocity $\dot{x%
}^{\mu }\left( \tau \right) $ and acceleration $\ddot{x}^{\mu }\left( \tau
\right) $. These remaining averages, which involve only the dynamical
variables of the event evolution, can be understood as artifacts of the
suppression of the runaway solutions. On the other hand, the combinations of
dynamical variables in these expressions derive from the form of the
radiation field (\ref{self-force}) emitted by the event, and therefore the
integrations may be seen as qualitatively analogous to the statistical
synchronization expressed in (\ref{pMm}) between the event and the field it
induces.

\end{document}